\documentclass[preprint2]{aastex6}
\usepackage{color}
\usepackage{times}
\usepackage{graphicx}
\usepackage{subfigure}
\usepackage{float}
\usepackage{txfonts}
\usepackage{epsfig}
\usepackage{epstopdf}
\usepackage{CJK}
\usepackage{natbib} 
\renewcommand{\baselinestretch}{0.9}

\def\tco{$^{13}$CO}
\def\ceo{C$^{18}$O}
\def\deg{^{\circ}}
\def\kms{km s$^{-1}$}
\renewcommand{\arraystretch}{2.0}

\newcommand{\methanol}{CH$_3$OH}
\newcommand{\ace}{CH$_3$CN}
\newcommand{\msun}{$M_\odot$}
\newcommand{\lsun}{$L_\odot$}
\newcommand{\hcop}{HCO$^+$}
\newcommand{\htco}{H$_2$CO}
\newcommand{\hii}{H~\textsc{ii}}
\slugcomment{Draft Version, \today}
\shorttitle{G22: a hub-filament cloud}
\shortauthors{Yuan et al.}

\begin{document}
\title{High-mass Star Formation through Filamentary Collapse 
	and Clump-fed Accretion in G22}
\author{Jinghua Yuan\altaffilmark{1}, Jin-Zeng Li\altaffilmark{1},
		Yuefang Wu\altaffilmark{2}, Simon P. Ellingsen\altaffilmark{3}, 
		Christian Henkel\altaffilmark{4,5}, Ke Wang\altaffilmark{6},
		Tie Liu\altaffilmark{7,8}, \\
		Hong-Li Liu\altaffilmark{9},  
		Annie Zavagno\altaffilmark{10},
		Zhiyuan Ren\altaffilmark{1}, Ya-Fang Huang\altaffilmark{1}
	    }
\affil{$^1$National Astronomical Observatories, Chinese Academy of Sciences, 
	20A Datun Road, Chaoyang District, \\
	Beijing 100012, China; jhyuan@nao.cas.cn;\\
    $^2$Department of Astronomy, Peking University, 100871 Beijing, China;\\
    $^3$School of Physical Sciences, University of Tasmania, Hobart, 
    Tasmania, Australia; \\
    $^4$Max-Planck-Institut f\"{u}r, Radioastronomie, Auf dem H\"{u}gel 69, 
    53121 Bonn, Germany; \\
    $^5$Astronomy Department, Faculty of Science, King Abdulaziz University,
    PO Box 80203, 21589 Jeddah, Saudi Arabia; \\
    $^6$European Southern Observatory, Karl-Schwarzschild-Str. 2, 
    D-85748 Garching bei M\"{u}nchen, Germany; \\
    $^7$Korea Astronomy and Space Science Institute 776, Daedeokdae-ro, 
    Yuseong-gu, Daejeon 34055, Korea; \\
    $^8$East Asian Observatory, 660 N. A'oh\={o}k\={u} Place, 
    Hilo, Hawaii 96720-2700, USA; \\
    $^9$Department of Physics, The Chinese University of Hong Kong,
    Shatin, NT, Hong Kong SAR; \\
    $^{10}$Aix Marseille Univ, CNRS, LAM, 
    Laboratoire d'Astrophysique de Marseille, 
    Marseille, France
	}

\begin{abstract}
	
	How mass is accumulated from cloud-scale down to individual 
	stars is a key open question in understanding 
	high-mass star formation. Here, we present
	the mass accumulation process in a hub-filament cloud G22
	which is composed of four supercritical filaments. Velocity
	gradients detected along three filaments indicate that
	they are collapsing with a total mass infall rate of
	about {440} \msun~Myr$^{-1}$, suggesting the hub mass would
	be doubled in {six} free-fall times, adding up to $ \sim2 $ Myr. 
	A fraction of the masses in the central 
	clumps C1 and C2 can be accounted for through 
	large-scale filamentary collapse. Ubiquitous blue profiles
	in \hcop~$ (3-2) $ and \tco~$ (3-2) $ spectra suggest
	a clump-scale collapse scenario in the most massive
	and densest clump C1. The estimated infall velocity and mass infall
	rate are {0.31} \kms~and $ 7.2 \times10^{-4} $ \msun~yr$^{-1}$,
	respectively. In clump C1, a hot molecular core (SMA1) is 
	revealed by the SMA observations and an outflow-driving 
	high-mass protostar is located at the center of SMA1. 
	The mass of the protostar is estimated to be $ 11-15 $ \msun~and it is
	still growing with an accretion rate of $ 7\times10^{-5} $ \msun~yr$^{-1}$.
	The coexistent infall in filaments, clump C1, and the central hot core
	in G22
	suggests that pre-assembled 
	mass reservoirs (i.e., high-mass starless cores) may not be 
	required to form high-mass stars. In the course of high-mass
	star formation, the central protostar, the core,
	and the clump can simultaneously grow in mass via core-fed/disk accretion,
	clump-fed accretion, and filamentary/cloud collapse.

\end{abstract}

\section{Introduction}

	The process of mass accumulation in high-mass star formation has remained 
	elusive for decades \citep{2014prpl.conf..149T,2017arXiv170600118M}.
	In the turbulent core model, which is one of the two extensively
	debated scenarios, the final stellar mass is pre-assembled in the 
	collapsing high-mass core \citep{2003ApJ...585..850M}. 
	Otherwise, the competitive accretion model
	allows the mass reservoir, often referred to as the star-forming core, to keep
	growing in the course of high-mass star formation \citep{2001MNRAS.323..785B}.
	
	Prevalent filaments detected in our Galaxy are well-established 
	as some of the main sites for star formation 
	\citep{2010A&A...518L.102A,2014prpl.conf...27A,
		2010A&A...518L.100M,2014MNRAS.439.3275W,
		2015MNRAS.450.4043W,2016ApJS..226....9W}. 
	Specifically, hub-filament systems
	are frequently reported forming high-mass
	stars \citep[e.g.,][]{2012A&A...543L...3H,2012ApJ...745...61L,
		2016ApJ...824...31L,2013A&A...555A.112P}. Converging flows 
	detected in several hub-filament systems
	channel gas to the junctions where star formation is often most active
	\citep[e.g.,][]{2013ApJ...766..115K,2014A&A...561A..83P,2016ApJ...824...31L}.
	However, how gas flows help individual star-forming cores grow in mass
	is still poorly understood.
	
	High-mass clumps ($ 0.3-1 $ pc) are the objects that fragment into
	dense cores ($ <0.3 $ pc) which subsequently contract to
	form individual or bound systems of protostars. 
	Clump-scale global collapse has been reported in case studies 
	\citep{2010A&A...520A..49S,2011A&A...527A.135C,2012MNRAS.422.1098R,
		2013MNRAS.436.1335L,2013A&A...555A.112P}, based on which 
	``clump-fed accretion'' has been proposed as an alternative
	high-mass star formation mode \citep{2017arXiv170309839T,2017arXiv170600118M}. 
	It still remains unknown, how significant such 
	clump-scale collapse may be in sustaining mass growth 
	in individual cores where high-mass protostars 
	have already formed.
	
	Located in the inner Galactic Plane with $ l=22.04\deg$ and $ b=0.2\deg $,
	the G22 cloud contains ten \textit{Spitzer} infrared
	dark clouds (IRDCs) from \citet{2009A&A...505..405P}. These IRDCs 
	are mainly distributed in a hub-filament system. Several young stellar 
	objects \citep{2008AJ....136.2413R}, 
	an extended green object \citep[EGO G022.04+00.22,][]{,2008AJ....136.2391C}, 
	and methanol maser emission \citep{2009ApJ...702.1615C} detected
	in G22 indicate active high-mass star formation therein. These features
	make G22 an ideal laboratory to investigate how mass is accumulated in 
	the high-mass star formation process, from cloud-scale 
	down to individual cores. Using
	the Bayesian Distance Calculator developed by \citet{2016ApJ...823...77R},
	the distance to G22 is estimated to be 
	$ 3.51\pm0.28 $ kpc based on its systemic velocity of 50.9 \kms. 
	
	In this work, we carry out an extensive study of G22 based on 
	observations with varying angular resolutions to 
	reveal a promising high-mass star formation scenario 
	where the central high-mass protostar, the core, 
	and the clump grow in mass simultaneously.
	The data used in this work are described in Section \ref{sec-data}. 
	We present results about the large scale cloud and the 
	SMA observations in Sections \ref{sect-large-cloud} 
	and \ref{sec-results-sma}. More in-depth discussions and 
	a summary of the findings are given in Sections 
	\ref{sec-discussions} and \ref{sec-summary}.

\section{Data} \label{sec-data}

	\subsection{Galactic Plane Survey Data}

	We extracted multi-wavelength data covering the entire cloud 
	from the GLIMPSE, MIPSGAL, Hi-GAL and ATLASGAL surveys.
	Using the IRAC instrument on board the \textit{Spitzer Space 
	Telescope} \citep{2004ApJS..154....1W}, the Galactic Legacy Infrared 
	Mid-Plane Survey Extraordinaire
	\citep[GLIMPSE,][]{2003PASP..115..953B,2009PASP..121..213C} project
	surveyed the inner $130\deg$ of the Galactic Plane at 3.6, 4.5, 
	5.8, and 8.0 \micron~with $5\sigma$ sensitivities of 0.2, 0.2, 
	0.4, and 0.4 mJy, respectively. The Galactic Plane Survey using 
	the Multiband Infrared Photometer on \textit{Spitzer}
	\citep[MIPSGAL,][]{2009ApJ...702.1615C} mapped an area comparable
	to GLIMPSE at longer infrared wavelengths. Version 3.0 of the MIPSGAL
	data provides mosaics from only the 24 \micron~band with 
	an angular resolution of 6\arcsec~and a $ 5\sigma $ sensitivity of 
	about 1.7 mJy. As a key project of the \textit{Herschel Space
	Observatory}, the \textit{Herschel} Infrared Galactic Plane Survey 
	\citep[Hi-GAL,][]{2010PASP..122..314M} mapped the entire Galactic plane
	with nominal $ |b|<1\deg $ (following the Galactic warp) at 
	70, 160, 250, 350, and 500 \micron~with angular resolutions 
	of 10$.\!\!\arcsec$2, 13$.\!\!\arcsec$5, 18$.\!\!\arcsec$1, 
	25$.\!\!\arcsec$0, and 36$.\!\!\arcsec$4. Using the LABOCA camera,
	the \textit{APEX} Telescope Large Areas Survey of the Galaxy 
	\citep[ATLASGAL][]{2009A&A...504..415S} mapped 420 square degrees of
	the Galactic plane between $ -80\deg<l<60\deg $ at 870 \micron~
	with a 19$.\!\!\arcsec$2 angular resolution. Cutouts of 
	$8\arcmin\times8\arcmin$ were retrieved from the above 
	surveys to reveal the large scale
	morphology and global properties of the entire cloud.
	Note that we used the ATLASGAL-\textit{Planck} combined
	data to better recover large scale features at 870 
	\micron~\citep{2016A&A...585A.104C}.
	
	\subsection{Single Dish Molecular Data}
	
	From the FCRAO Galactic Ring Survey \citep[GRS,][]{2006ApJS..163..145J}
	we retrieved \tco~$J=1-0$ line data of the cloud. The GRS \tco~
	data have velocity and angular resolutions of 0.21 \kms~and 46\arcsec.
	The main beam efficiency ($ \eta_\mathrm{mb} $) and the typical
	rms sensitivity are 0.48 and 0.13 K, respectively.
	
	From the JCMT data archive, we extracted \tco~$ (3-2) $,
	\ceo~$ (3-2) $,
	and \hcop~$ (3-2) $ data which cover only the central 
	$ \sim1\arcmin\times1\arcmin $. The
	\tco~$ (3-2) $ and
	\ceo~$ (3-2) $ were simultaneously observed in August of 2007
	with an angular resolution 
	of 15\arcsec, a velocity resolution of 0.055 \kms, 
	and a rms noise of about 1.20 K. Observed in May of 2008, the \hcop~$ (3-2) $ data
	have an angular resolution of 20\arcsec, a velocity resolution of 0.55 \kms, 
	and a rms noise of about 0.18 K.
	
	\subsection{SMA Observations}

    Observations towards the central massive clump C1 
    using the Submillimeter Array (SMA) at 1.3 mm were performed on 
    2010 August ${\rm 10^{th}}$ in the compact-north 
    configuration with eight antennae used. {The
    data are publicly available in the SMA data archive
    (PI: Claudia Cyganowski).} The $2\times4$ GHz correlator 
    was tuned to cover the frequency range 216.8--220.8 GHz in the lower sideband (LSB) 
    and 218.8--232.8 GHz in the upper sideband (USB). The 4 GHz bandwidth in 
    each sideband is allocated to 48 spectral windows, each of which consists
    of 128 channels with an uniform width of 
    812.5 kHz ($\sim$1.1 km s$^{-1}$). The projected baseline lengths ranged from 8 to 
    105 k$\lambda$, which indicates that the SMA observations are
    insensitive to smoothed structures larger than ~$20\arcsec$.

    3c454.3, J1733-130/J1911-201, and Callisto (Jupiter's moon) were observed 
    as bandpass, gain and flux density calibrators.
    The data were reduced and imaged 
    using MIRIAD \citep{1995ASPC...77..433S}. GILDAS \citep{2005sf2a.conf..721P}, 
    APLpy\footnotemark[1] and Astropy \citep{2013A&A...558A..33A} packages
    were used for data visualization and analysis.  
    The continuum data were constructed 
    using line free channels from both sidebands, which resulted in a 
    synthesized beam of $\mathrm{2.\!\!\arcsec95\times1.\!\!\arcsec92\ (PA=68.4\deg)}$ and 
    a rms noise of  3.2 mJy beam$^{-1}$. For the 
    line data, the continuum emission was subtracted by modeling an order one 
    polynomial. We also performed self-calibration to the continuum data 
    and applied the solutions to the line data. The typical noise in a 
    single channel of the line data is about 80 mJy beam$^{-1}$. 
    
    \footnotetext[1]{APLpy is an open-source plotting package for 
    	Python hosted at \url{http://aplpy.github.com}.}
    
\section{A Hub-Filament Cloud}\label{sect-large-cloud}

	\begin{deluxetable}{ccccc}
		\tablecaption{Properties of filaments \label{tb-filaments}}
		\tablewidth{0pt}
		\tablehead{ \colhead{Name} & \colhead{Length} & \colhead{Mass} & \colhead{$ M_\mathrm{line} $} &
			\colhead{$ N_\mathrm{H_2} $\tablenotemark{a}}  \\
			& \colhead{(pc)} & \colhead{(\msun)} & \colhead{(\msun~pc$^{-1}$)} & \colhead{($ 10^{21} $ cm$^{-2}$)} }
		\startdata
		F1 & 2.91 & 357 & 123 & 4.59\\
		F2 & 2.62 & 575 & 220 & 7.28 \\
		F3 & 3.02 & 485 & 161 & 7.72 \\
		F4 & 1.88 & 102 &  54 & 3.07 \\
		\enddata 
		\tablenotetext{a}{Averaged column density.}
	\end{deluxetable}
        
    \begin{figure*}[htb]
    	\centering
    	\includegraphics[width=0.98\textwidth]{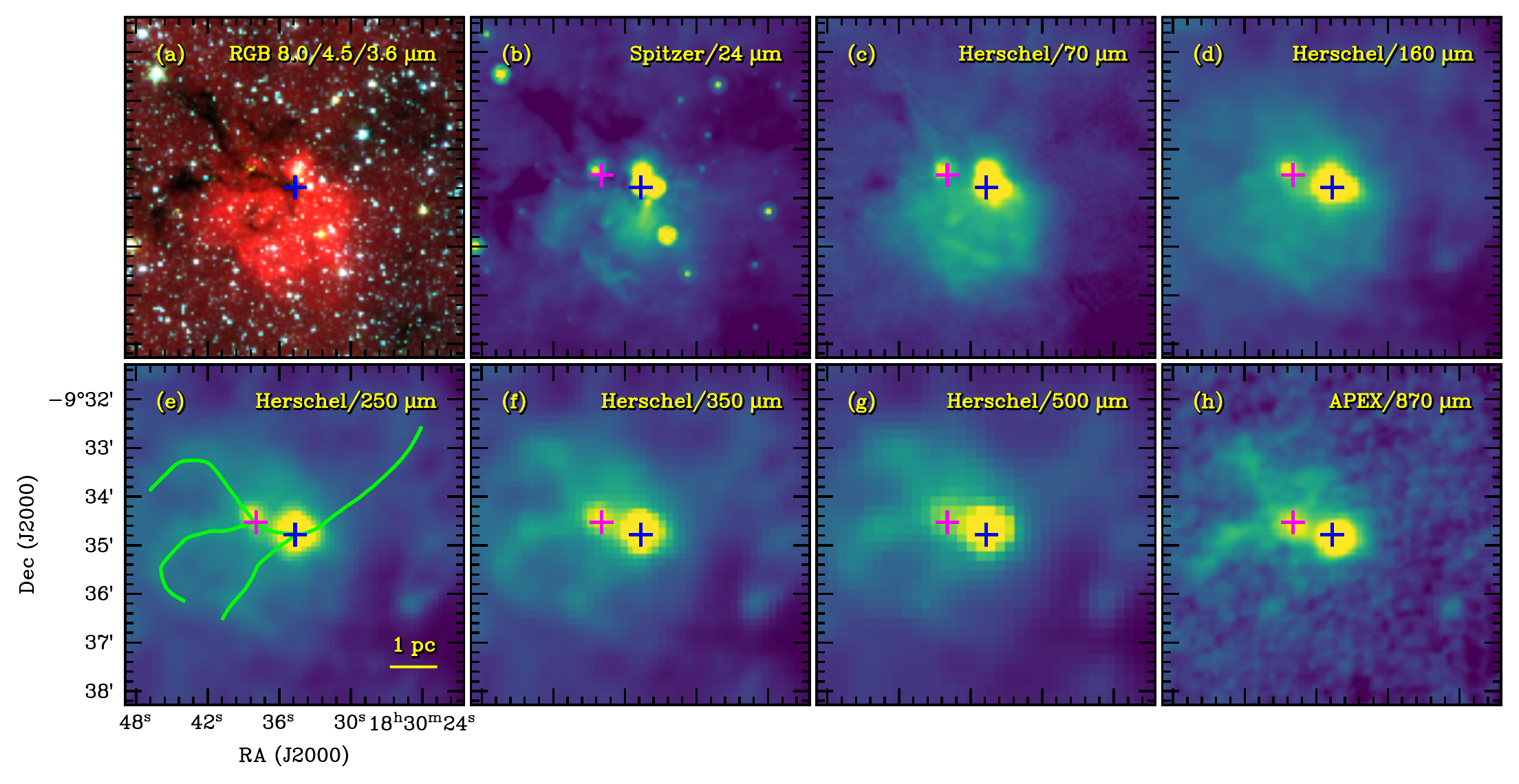}
    	\caption{Morphology of G022.04+0.22 in multi-wavelength bands.
    		The blue and magenta crosses mark the most massive clumps 
    		C1 and C2 (see Figure \ref{fig-Nh2-Tdust} and
    		Section \ref{sect-largeDust}) in the hub region.
    		The skeletons of the four filaments are also shown in (e).
    		{The Hi-GAL images shown here are the unfiltered ones.}
    		\label{fig-mul-morph}}
    \end{figure*}

	Cutouts at wavelengths from 3.6 to 870 \micron~of the G22 cloud are shown in
	Figure \ref{fig-mul-morph}. Extinction features, which correspond 
	to 10 infrared dark clouds identified by \citet{2009A&A...505..405P}, 
	dominate in the north at 8.0 \micron~and can be
	partly seen at 24 and 70 \micron. Extended emission in 
	the south at 8.0, 24, and 70 \micron~originates from an infrared dust bubble, 
	MWP1G022027+002159 \citep{2012MNRAS.424.2442S}.
	This infrared dust bubble is slightly farther from us but physically 
	interacting with G22 (see Section \ref{sec-discussion-bubble}).
	
	As shown in Figure \ref{fig-mul-morph} (b)--(h), two clumps (C1 and C2, see 
	bellow) dominate emission at wavelengths longer than 24 \micron.
	These two clumps also represent the most active star-forming regions in
	G22. Detected at far-IR and sub-mm wavelengths, 
	several filaments converge at clumps C1 and C2.
	
	\subsection{Dust Properties}\label{sect-largeDust}
	
		\begin{figure*}[htb]
		\centering
		\includegraphics[width=1.0\textwidth]{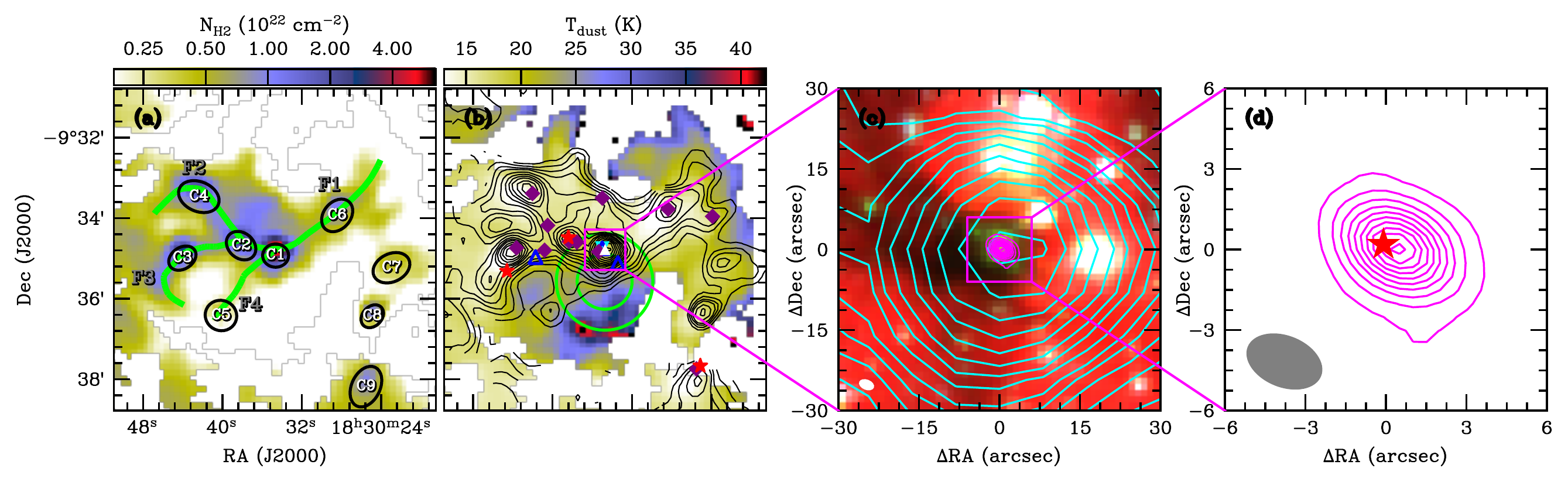}
		\caption{{\it(a)} Column density map from SED fitting. 
			The eight clumps, which have been identified based on this map and 
			designated to be C1 to C8, are shown as open ellipses. 
			Also labeled are the visually identified filaments F1 to F4.
			\textit{(b)} Dust temperature map from SED fitting with column density
			overlaid as contours in levels of  $(0.2,~0.3,~0.4,~0.5,~0.7,~
			0.9,~1.1,~1.3,~1.6,~1.9,~
			2.3,~2.8,~3.4,~4.1,~5.0,~6.1)\times10^{22}$ cm$^{-1}$.
			Infrared dark clouds from \citet{2009A&A...505..405P} are marked as purple
			diamonds. The red stars represent young stellar object candidates 
			from \citet{2008AJ....136.2413R}. EGO G022.04+0.22 is shown as a filled triangle.
			The three blue open triangles represent molecular hydrogen emission objects
			from \citet{2012MNRAS.421.3257I}. The cyan filled inverted triangle labels 
			IRAS 18278-0936. The two green ellipses delineate the inner and outer rings
			of the infrared bubble MWP1G022027+002159 \citep{2012MNRAS.424.2442S}.
			{\it(c)} H$_{2}$ column density from SED fits
			(cyan contours) 
			and SMA 1.3 mm continuum emission (magenta contours) 
			overlaid on a IRAC three
			color image with emission at 8.0, 4.5, and 3.6 \micron~rendered in
			red, green, and blue, respectively. Cyan contours represent
			H$_{2}$ column densities of $(0.7,~0.9,~1.1,~1.3,~1.6,~1.9,~
			2.3,~2.8,~3.4,~4.1,~5.0,~6.1)\times10^{22}$ cm$^{-1}$.
			Magenta contour levels at the center (better visible in (d)) start 
			from $10\sigma$ and increase with a step of $10\sigma$, 
			where $1\sigma=3.2$ mJy beam$^{-1}$. 
			{\it(d)} A close-up view of SMA 1.3 mm continuum. The {single core} is designated as SMA1. 
			A filled star represents 
			the MIR source SSTGLMC G022.0387+00.2222 (MIR1) from the GLIMPSE survey. 
			\label{fig-Nh2-Tdust}}
	\end{figure*}
	
	Column density and dust temperature maps have been 
	obtained via fitting the spectral
	energy distribution (SED) to the multi-bands data. 
	To reach an intermediate angular resolution
	we only used data at 160, 250, 350, and 870 \micron~to 
	perform the fitting. 
	The images were convolved to a common angular 
	resolution of 25$\arcsec$ which
	is essentially the poorest resolution of the 
	considered wavelengths. {The convolution kernels 
		of \citet{2011PASP..123.1218A} were used to take the 
		different instrumental responding functions into account.
	For the ATLASGAL images, any uniform astronomical 
	signal on spatial scales larger than 2.\arcmin5 has 
	been filtered 
	out together with atmospheric emission during the data 
	reduction \citep{2009A&A...504..415S}. The filtering of 
	Hi-GAL images was performed using the 
	CUPID-findback algorithm of the Starlink suite. 
	For further details on the algorithm please see the online 
	document for findback\footnotemark[2]. Following \citet{2017ApJS..231...11Y},
    we run the algorithm iteratively using a common filtering box
	of 2.\arcmin5 to get stable background images which 
	were subtracted from the post-convolution data for Hi-GAL 
	bands to remove large-scale structures.}

	\footnotetext[2]{http://starlink.eao.hawaii.edu/starlink/findback.html}
	
	
	Intensity as a function of wavelength for each pixel was obtained from the
	smoothed and background-removed far-IR to sub-mm image data and modeled 
	to a modified blackbody. 
	\begin{equation}\label{eq-gb}
	I_\nu=B_\nu(T) 
		  \mu_\mathrm{H_2}m_\mathrm{H}\kappa_\nu N_\mathrm{H_2}/R_\mathrm{gd}
	\end{equation}
	where 
	\begin{equation}
	\kappa_\nu=\kappa_\mathrm{600}\left(\frac{\nu}{600~\mathrm{GHz}}\right)^\beta~\mathrm{cm^2g^{-1}}.
	\end{equation} 
	Here, $\mu_\mathrm{H_2}=2.8$ 
	is the mean molecular weight adopted from \citet{2008A&A...487..993K},
	$m_\mathrm{H}$ is the mass of a 
	hydrogen atom, $N_\mathrm{H_2}$ is the H$_2$ column density, $R_\mathrm{gd}=100$ is 
	the gas to dust ratio, $\kappa_\mathrm{600}=5.0~\mathrm{cm^2g^{-1}}$ is the dust opacity
	at 600 GHz for
	coagulated grains with thin ice mantles  
	\citep{1994A&A...291..943O}.
	The dust emissivity index $ \beta $ was
	fixed to 1.75 \citep{1994A&A...291..943O,2014A&A...564A..45P}. 
	Free parameters in this model are the dust temperature $ N_\mathrm{H_2} $
	and column density $ T_\mathrm{dust} $. 
	
	The resultant column density and dust temperature maps are presented in
	Figure \ref{fig-Nh2-Tdust}. We note that the dust temperatures will be
	underestimated as the modeling only considered emission at wavelengths 
	not shorter than 160 \micron, while some regions significantly emit at 70 \micron.
	Uncertainties in $ N_\mathrm{H_2} $
	and $ T_\mathrm{dust} $ can also originate from the dust opacity, which is 
	subject to a factor of two uncertainty \citep{1994A&A...291..943O}. 
	The dust emissivity index ($ \beta $)
	can also largely influence the resultant parameters. An increase of 0.5 for $ \beta $
	would lead to a $ 2\%-35\% $ increase for $ N_\mathrm{H_2} $ and a $ 5\%-18\% $ 
	decrease for $ T_\mathrm{dust} $, and $ N_\mathrm{H_2} $ would decrease by 
	$ 10\%-35\% $ and  $ T_\mathrm{dust} $ might increase by $ 13\%-28\% $ if $ \beta $
	decreases by 0.5 \citep{2017ApJS..231...11Y}.
	
	\begin{deluxetable*}{ccccccccccccc}
		\tablecaption{Properties of the identified dense clumps \label{tb-clumps}}
		\tablewidth{0pt}
		\tablehead{ \colhead{Clulmp} & \colhead{R.A.} & \colhead{Dec.} & 
			\colhead{$ \Theta_\mathrm{maj} $} & \colhead{$ \Theta_\mathrm{min} $} &
			\colhead{PA} & \colhead{FWHM}\tablenotemark{a} & \colhead{$ r_\mathrm{eq} $\tablenotemark{b}} & 
			\colhead{$ T_\mathrm{dust} $} &
			\colhead{$ N_\mathrm{H_2} $} & \colhead{$ M_\mathrm{cl} $} &
			\colhead{$ \Sigma_\mathrm{mass} $} &\colhead{$ n_\mathrm{H_2} $}\\
			& & & \colhead{(\arcsec)} & \colhead{(\arcsec)} & 
			\colhead{($ \deg $)} & \colhead{(\arcsec)} & \colhead{(pc)} & 
			\colhead{(K)} & \colhead{($ 10^{22} $ cm$^{-2}$)} & \colhead{(\msun)} & 
			\colhead{(g cm$^{-2}$)} & \colhead{($ 10^{4} $ cm$^{-3}$)}
		}
		\startdata
		C1 & 18$^\mathrm{h}$30$^\mathrm{m}$34.64$^\mathrm{s}$ & -9$^\mathrm{d}$34$^\mathrm{m}$55.9$^\mathrm{s}$  &  38.8 &  34.6 &  90.3 &  36.5 & 0.39  & 20.9 & 6.55 &  590  & 0.26 & 3.52 \\ 
		C2 & 18$^\mathrm{h}$30$^\mathrm{m}$38.12$^\mathrm{s}$ & -9$^\mathrm{d}$34$^\mathrm{m}$41.1$^\mathrm{s}$  &  48.1 &  37.8 &  49.2 &  42.3 & 0.49  & 18.4 & 2.39 &  319  & 0.09 & 0.91 \\ 
		C3 & 18$^\mathrm{h}$30$^\mathrm{m}$44.04$^\mathrm{s}$ & -9$^\mathrm{d}$34$^\mathrm{m}$59.7$^\mathrm{s}$  &  43.9 &  32.1 & 123.4 &  36.8 & 0.39  & 13.8 & 2.68 &  270  & 0.12 & 1.56 \\ 
		C4 & 18$^\mathrm{h}$30$^\mathrm{m}$42.38$^\mathrm{s}$ & -9$^\mathrm{d}$33$^\mathrm{m}$28.0$^\mathrm{s}$  &  62.9 &  44.3 &  67.2 &  52.4 & 0.67  & 14.8 & 2.17 &  409  & 0.06 & 0.48 \\ 
		C5 & 18$^\mathrm{h}$30$^\mathrm{m}$40.15$^\mathrm{s}$ & -9$^\mathrm{d}$36$^\mathrm{m}$25.2$^\mathrm{s}$  &  47.4 &  45.0 &  66.3 &  46.2 & 0.56  & 23.5 & 0.41 &   62  & 0.01 & 0.12 \\ 
		C6 & 18$^\mathrm{h}$30$^\mathrm{m}$28.44$^\mathrm{s}$ & -9$^\mathrm{d}$33$^\mathrm{m}$56.0$^\mathrm{s}$  &  53.9 &  40.2 & 139.6 &  46.1 & 0.56  & 14.0 & 1.11 &  174  & 0.04 & 0.34 \\ 
		C7 & 18$^\mathrm{h}$30$^\mathrm{m}$22.97$^\mathrm{s}$ & -9$^\mathrm{d}$35$^\mathrm{m}$14.0$^\mathrm{s}$  &  57.6 &  41.7 & 116.0 &  48.5 & 0.60  & 19.3 & 0.54 &   96  & 0.02 & 0.15 \\ 
		C8 & 18$^\mathrm{h}$30$^\mathrm{m}$24.90$^\mathrm{s}$ & -9$^\mathrm{d}$36$^\mathrm{m}$26.3$^\mathrm{s}$  &  38.0 &  29.0 & 137.0 &  32.3 & 0.30  & 15.9 & 0.98 &   79  & 0.06 & 1.03 \\ 
		C9 & 18$^\mathrm{h}$30$^\mathrm{m}$25.54$^\mathrm{s}$ & -9$^\mathrm{d}$38$^\mathrm{m}$11.2$^\mathrm{s}$  &  63.9 &  42.5 & 154.6 &  51.5 & 0.65  & 15.0 & 1.00 &  190  & 0.03 & 0.24 \\
		\enddata
		\tablenotetext{a}{Full width at half maximum: $ \mathrm{FWHM} = 
			 \sqrt{\Theta_\mathrm{maj}\Theta_\mathrm{min}}. $}
		\tablenotetext{b}{Equivalent radius: 
			$r_\mathrm{eq} = d\sqrt{\Theta_\mathrm{maj}\Theta_\mathrm{min}-
				\theta_\mathrm{beam}^2}/\sqrt{2\ln2}$, where $ \theta_\mathrm{beam} = 25\arcsec $.}
	\end{deluxetable*}
	
	From the $ N_\mathrm{H_2} $ map, we have visually 
	identified four filamentary structures 
	which intersect at the center to
	constitute a hub-filament system. These four filaments
	are named as F1 to F4 and listed in 
	Table \ref{tb-filaments} together with some physical parameters.
	Their loci are overlaid in Figure \ref{fig-Nh2-Tdust} (a). 
	{Nine $N_\mathrm{H_2}$ peaks were identified 
	as clumps and fitted to 2D Gaussian profiles using the 2D Fitting Tool
	of the CASA viewer.}
	Their FK5 coordinates, 
	and major and minor axes ($\Theta_\mathrm{maj}$ and 
	$\Theta_\mathrm{min}$) are given in Table 
	\ref{tb-clumps}. These clumps are designated as
	C1 to C9 and delineated as open ellipses in Figure
	\ref{fig-Nh2-Tdust} (a). Interestingly, six of
	the nine clumps are located inside filaments with the most massive
	one (C1) in the intersecting hub.

	
	Clump masses were estimated via integrating the column densities
	in the Gaussian ellipses. Source-averaged
	$\mathrm{H_2}$ number densities and mass surface densities 
	were calculated assuming a spherical morphology with a
	constant density profile and an equivalent radius of
	{$r_\mathrm{eq} = d\sqrt{\Theta_\mathrm{maj}\Theta_\mathrm{min}-
		\theta_\mathrm{beam}^2}/\sqrt{2\ln2}$, 
	where $ \theta_\mathrm{beam} = 25\arcsec $.}
	The resultant equivalent radii, clump masses, $\mathrm{H_2}$ number densities,
	and source-averaged surface densities are given in Table 
	\ref{tb-clumps} together with peak 
	$\mathrm{H_2}$ column densities and source-averaged dust temperatures. These eight clumps
	have {masses ranging from 62 to 590 \msun, radii ranging from 0.30 to
		0.67 pc. 
	The total mass of all clumps is 2189 \msun, holding about
	47\% of the cloud mass (5943 \msun)}. 
	
	\subsection{Molecular Emission}

	\begin{figure}
		\centering
		\includegraphics[width=0.48\textwidth]{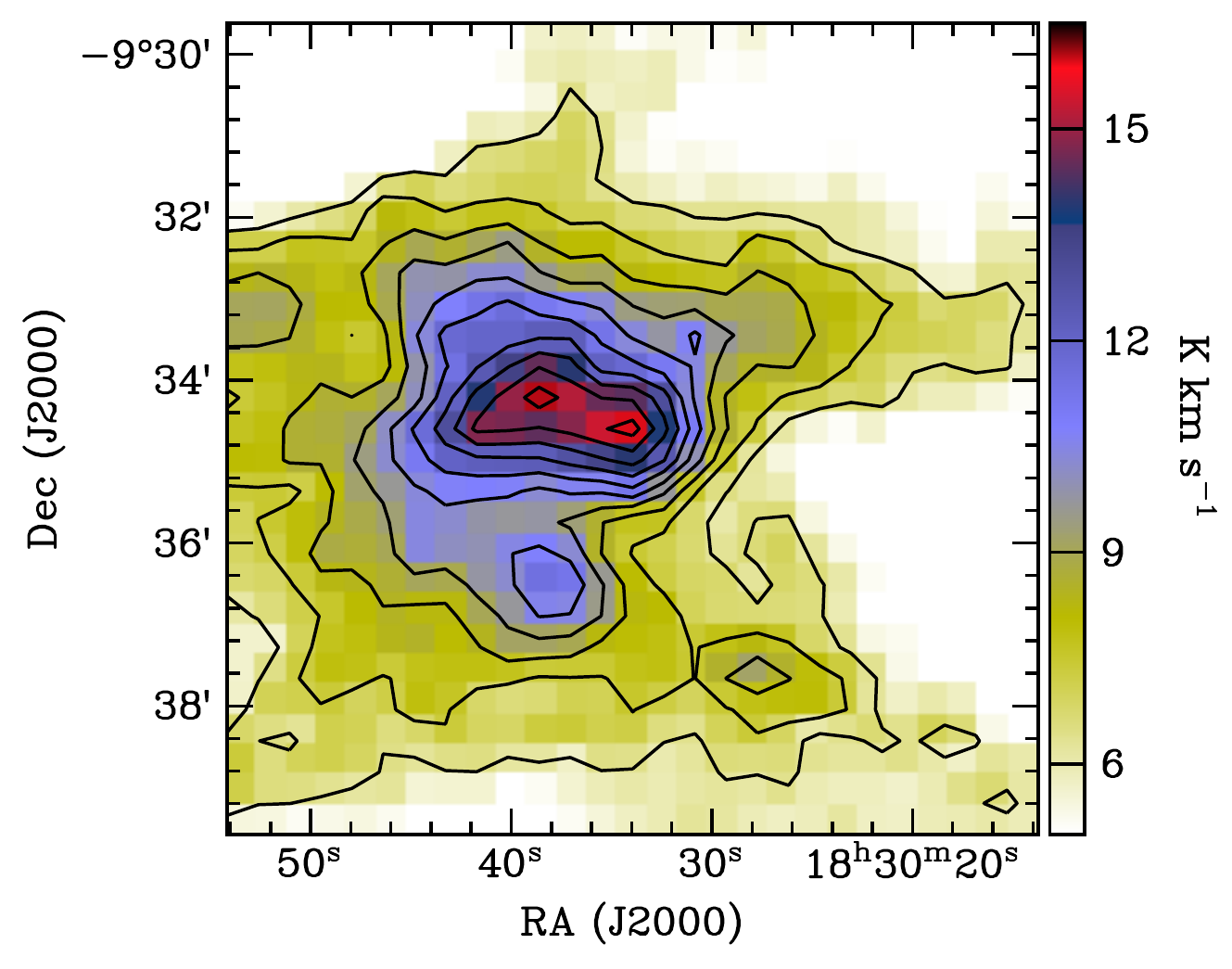}
		\caption{Velocity-integrated intensity map of \tco~$(1-0)$. 
			The velocity interval for integration is [46, 59] \kms. 
			The contours start from 6.5 K \kms~and increase with
			a step of 1.0 K \kms.
			\label{fig-GRS-m0}}
	\end{figure}

	The velocity-integrated intensity map of \tco~$(1-0)$ is shown in
	Figure \ref{fig-GRS-m0}. Although filamentary features are smoothed out
	due to the poor angular resolution, the strongest emission is still
	spatially coincident with the hub region.

	Under local thermodaynamic equilibrium (LTE) conditions, 
	the column density of a linear molecule can be expressed as
	\begin{eqnarray}
	N &=&\frac{3k}{8\pi^3\nu\mu^2S}\frac{Q_\mathrm{rot}}{g_{J+1}}
	\frac{\mathrm{exp}\left(\frac{E_\mathrm{up}}{k T_\mathrm{ex}}\right)}{J(T_\mathrm{ex})} \nonumber \\
	& &\times\frac{1}{J(T_\mathrm{ex})-J(T_\mathrm{bg})}\frac{\tau}{1-\mathrm{exp}(-\tau)}\int T_\mathrm{r}~dv \label{eq-lte-density}.
	\end{eqnarray}
	Here, $\mu$ is the permanent dipole moment. For the $1-0$ transition, 
	the line strength $S = \frac{J+1}{2J+3} = \frac{1}{3}$, the degeneracy 
	$g_\mathrm{J+1} = 2J+3 = 3$, where $J$ is the rotational quantum 
	number of the lower state. We have followed \citet{1988JChPh..88..356M} 
	to write the partition function as $Q_\mathrm{rot} = \frac{k T}{hB}+\frac{1}{3}$ 
	where $B$ is the rotational constant. 
	
	In the calculation, a single excitation 
	temperature was assumed to be 20 K which is approximately the averaged 
	dust temperature of the cloud. The H$_2$ column density in each pixel 
	was obtained by assuming a $\mathrm{^{12}C/^{13}C=50}$ isotope
	ratio \citep[estimated using Equation 4 of][]{2013A&A...554A.103P}
	and a canonical CO abundance of $X_\mathrm{CO} = 10^{-4}$.
	The resultant column densities range from 
	$ 3.4\times10^{21} $ to $ 2.19\times10^{22} $ cm$^{-2}$ with a mean of
	$ 9.39\times10^{21} $ cm$^{-2}$. Integrating the whole density map resulted
	in a total mass of 6117 \msun. We note that \tco~$(1-0) $ is assumed to
	be optically thin and the resultant cloud mass might be lower limit.
%

\section{Results of SMA Observations}\label{sec-results-sma}

    \subsection{1.3 mm Continuum Emission}\label{sec:cont}


	A zoom-in IRAC three color image of C1, 
	the most massive clump in G22, is shown in
	Figure \ref{fig-Nh2-Tdust} (c) with $N_\mathrm{H_2}$ and the SMA 1.3 mm 
	continuum emission overlaid as cyan and magenta contours. 
	With the current SMA angular resolution and sensitivity, only one 1.3 mm continuum
	core has been detected at the center and designated as SMA1. 
	
	A close-up view of SMA1 is shown in Figure \ref{fig-Nh2-Tdust} (d).
    Two-dimensional Gaussian fitting of the 1.3 mm continuum 
    emission resulted in an ellipse with 
    a major axis of $3.\!\!\arcsec68$, a 
    minor axis of $2.\!\!\arcsec69$ and a position angle of 
    $56.\!\!^{\circ}87$. The deconvolved major and minor axes are 
    $2.\!\!\arcsec34$ and $1.\!\!\arcsec70$, 
    smaller than the synthesized beam (${2.\!\!\arcsec95\times1.\!\!\arcsec92}$).
    The equivalent angular size of the core is $1.\!\!\arcsec99$, 
    corresponding to a physical diameter of 
    0.034 pc ($\sim7000$ AU). The flux density of the core is about 562 mJy 
    with a peak intensity of 350 mJy beam$^{-1}$.

    Assuming the 1.3 mm continuum 
    emission is optically thin, the core 
    dust mass can be obtained via  
    $M=R_\mathrm{gd}S_\nu d^2/[\kappa_\nu B_\nu(T)]$, where $S_\nu$ is the 
    observed flux density, $d$ is the distance and 
    $\kappa_\nu=1.0$  cm$^2$ g$^{-1}$ is the dust 
    opacity \citep{1994A&A...291..943O}.  
    Using the rotational temperature of \methanol~($ \sim227 $ K, 
    see Section \ref{sec-temperature}) 
    as the dust temperature, 
    a core mass of $10.4$ $M_\sun$ 
    is reached. The source averaged H$_2$ number density
    is about $7.2\times10^{6}$ cm$^{-3}$. 
    The beam-averaged column density was estimated 
    to be about $1.55\times10^{23}$ cm$^{-2}$
    via $N_\mathrm{H_2} = 
    I_\nu R_\mathrm{gd}/[\kappa_\nu B_\nu(T)m_\mathrm{H}\mu_\mathrm{H_2}]$.

    \subsection{Line Emission}\label{sec-line}

    \begin{figure*}[htb]
      \centering
      \includegraphics[width=0.95\textwidth]{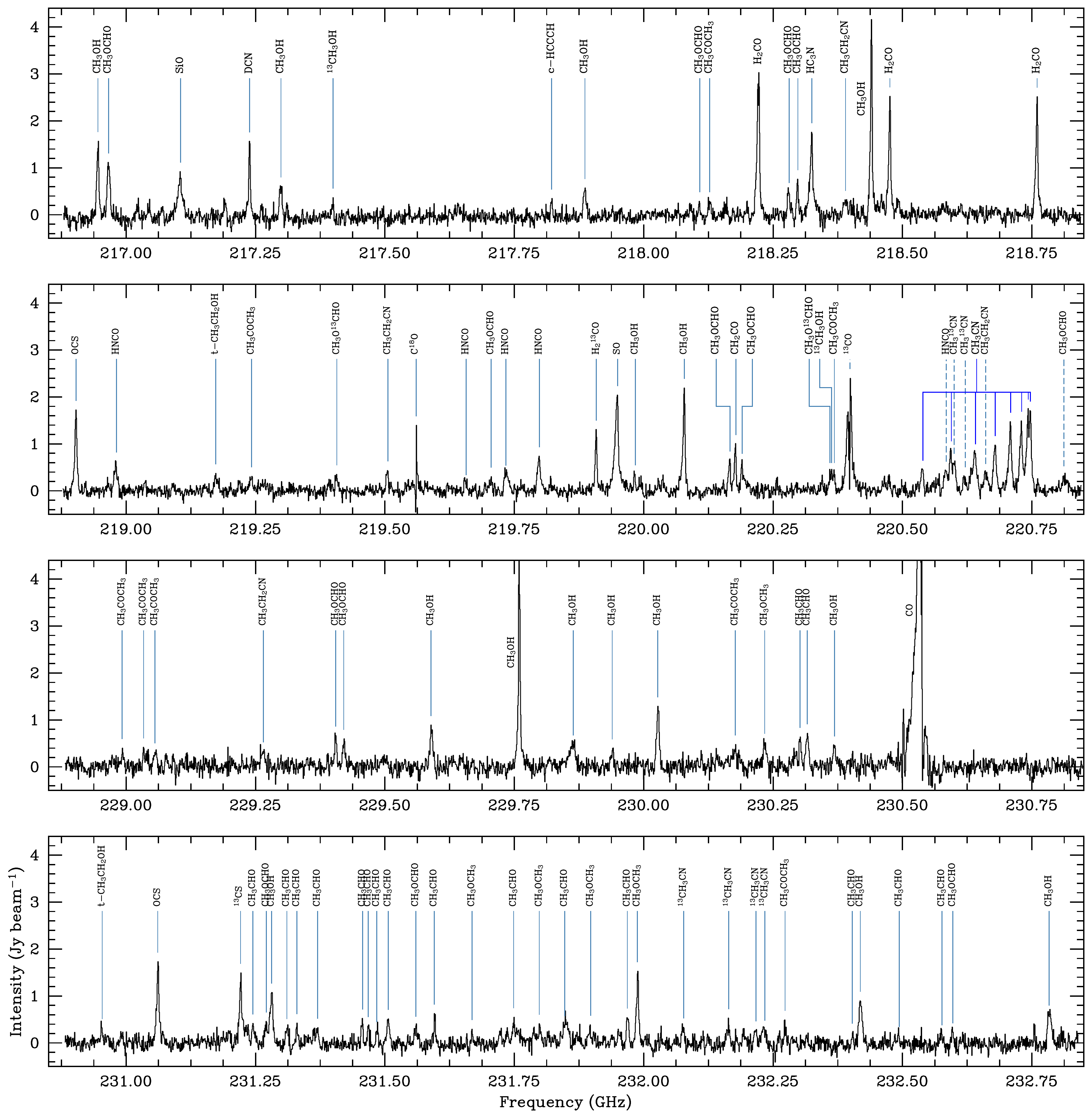}
      \caption{SMA LSB and USB spectra toward the 
      peak of SMA1. Lines with peaks no smaller than 3$\sigma$ (240 mJy beam$^{-1}$) 
      have been labeled. \label{fig-spec-all}}
    \end{figure*}


    Figure \ref{fig-spec-all} shows the full band spectra 
    towards the SMA1 peak. 
    Cross-checking the rest frequencies with the \textit{Splatalogue} 
    molecular database, we have identified 103 emission lines with
    peaks $>3\sigma$ where $1\sigma = 80$ mJy beam$^{-1}$. These transitions are
    from 19 species (26 isotopologues) including nitrogen-bearing species
    (e.g., HC$_3$N, \ace), complex organic molecules (e.g., CH$_3$CHO,
    CH$_3$OCH$_3$), and classical tracers of bulk motions (e.g., 
    CO, SiO). 
    
    We extracted spectra of all detected transitions towards the SMA1 
    peak and used the
    $ GILDAS/CLASS $ software to perform Gaussian profile fitting
    to get the line peak intensities, velocity centroids, linewidths,
    and velocity-integrated intensities. 
    The systemic velocity of SMA1 
    was estimated to be $50.3\pm0.1$ km s$^{-1}$ via 
    averaging line centers of transitions with $\mathrm{FWHM}\leq9$ km s$^{-1}$ 
    and $I_{\mathrm{peak}}\geq0.4$ Jy beam$^{-1}$.
    
    Inspection of velocity-integrated intensity maps shows that lines from 19 molecules 
    trace a single-core morphology in accord with
    the 1.3 mm continuum emission. 
    On the other hand, lines of CO, \tco, \ceo, 
    \htco, SO, and SiO reveal extended structures either from the surrounding
    envelope or entrained outflows (see Section \ref{sec-outflow}). Among the
    15 detected \methanol~lines, 13 trace a single-core spatially 
    coincident with the 1.3 mm core. The other two methanol transitions
    show anomalous emission in the vicinity (see Section \ref{sec-results-maser}).

    \subsection{Molecular Outflows}\label{sec-outflow}

    \begin{figure*}[htb]
        \centering
        \includegraphics[width=0.9\textwidth]{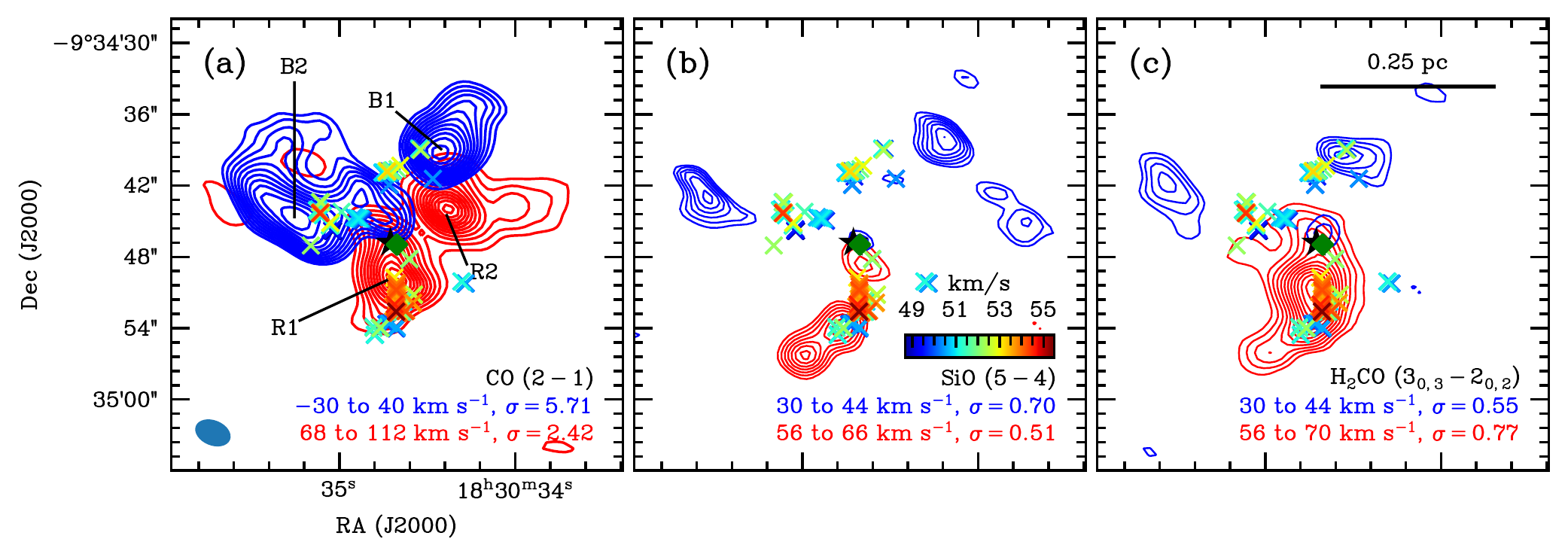}
        \caption{Outflows revealed by CO $ (2-1) $ \textit{(a)}, SiO $(5-4)$ \textit{(b)},
        	and H$_2$CO $ (3_{0,3}-2_{0,2}) $ \textit{(c)}. 
        	Velocity intervals and $ 1\sigma $ noise levels of the red and blue 
        	lobes for each line are labeled in the bottom-right corner of each panel.
        	For CO $ (2-1) $, the blue contours start from $ 3\sigma $ and increase 
        	by $ 2\sigma $, the blue contours start from $ 3\sigma $ and increase by
        	$ 1\sigma $. For SiO $ (5-4) $, the blue contours start from $ 3\sigma $ and increase 
        	by $ 1\sigma $, the blue contours start from $ 3\sigma $ and increase by
        	$ 0.5\sigma $. 
        	For H$_2$CO $ (3_{0,3}-2_{0,2}) $, the blue contours start from $ 3\sigma $ and increase 
        	by $ 1\sigma $, the blue contours start from $ 3\sigma $ and increase
        	$ 1\sigma $. 
        	The 44 GHz Class I methanol masers from \citet{2009ApJ...702.1615C}
        	are marked with crosses and color-coded according to their velocities.
        	The green diamond and the black star show the locations of the 1.3 mm core 
        	SMA1 and embedded protostar SSTGLMC G022.0387+00.2222 (MIR1).
        	\label{fig-outflow}}
    \end{figure*}

	\begin{deluxetable*}{cccccccc}
		\tablecaption{Outflow Parameters \label{tb-outflow}}
		\tablewidth{0pt}
		\tablehead{ \colhead{Lobe} & \colhead{$L_\mathrm{flow}$} & 
			\colhead{$v_\mathrm{char}$} & 
			\colhead{$M_\mathrm{out}$} & \colhead{$P_\mathrm{out}$} & 
			\colhead{$E_\mathrm{out}$} & \colhead{$t_\mathrm{dyn}$} & 
			\colhead{$\dot{M}_\mathrm{out}$} \\
			&  \colhead{(pc)} &\colhead{(\kms)} & 
			\colhead{($M_\sun$)} & \colhead{($M_\sun$ \kms)} & 
			\colhead{($M_\sun$ km$^2$ s$^{-2}$)} & 
			\colhead{(10$^4$ yr)} & \colhead{(10$^{-5}$ $M_\sun$ yr$^{-1}$)}
		}
		\startdata
		B1  & 0.26 &  22.47 &  0.57 & 12.96 & 182.94 & 1.12 & 5.16 \\ 
		B2  & 0.27 &  19.63 &  1.23 & 24.22 & 300.82 & 1.33 & 9.25 \\ 
		R1  & 0.13 &  27.43 &  0.37 & 10.21 & 158.28 & 0.47 & 7.95 \\ 
		R2  & 0.26 &  28.18 &  0.38 & 10.71 & 170.64 & 0.89 & 4.29 \\
		\hline
		Total &\nodata&\nodata&2.56&58.10&812.68&\nodata&26.65 \\
		\enddata
	\end{deluxetable*}

	In multi-object surveys, previous works have detected outflow indicators 
	towards EGO G022.04+0.22 (C1 in our work), including
	class I methanol masers \citep{2009ApJ...702.1615C,2011ApJS..196....9C}, 
	``red-skewed'' asymmetric line profiles of CO and \hcop~\citep{2013A&A...549A...5R},
	and broad SiO $ (5-4) $ emission \citep{2009ApJ...702.1615C}. The existence 
	of outflows in C1 is confirmed by the SMA observations presented in this work. 
	
	SMA CO $(2-1)$ shows very broad
	($>70$ \kms~with respect to the systemic velocity) wing emission. 
	Velocity-integrated intensities of the blue and red wings 
	(see Figure \ref{fig-outflow} (a)) reveal two blue-shifted and
	two red-shifted lobes (B1, B2, R1 and B2).
	These outflow lobes can be well traced back to the
	SMA1 peak. 
	{The multiple lobes of the outflow also could be 
	due to the presence of two, unresolved protostellar objects.}
	The outflows are also
	seen in emission from SiO and 
	\htco~(see Figure \ref{fig-outflow} (b) and (c)). 
	The GLIMPSE point source 
	SSTGLMC G022.0387+0.2222 (MIR1, hereafter) 
	which is an early-stage high-mass 
	protostar (see Section \ref{sec-hot-core}), associated 
	with SMA1 could be the driving source.

    Copious 44 GHz Class I \methanol~masers detected by 
    \citet{2009ApJ...702.1615C} are shown in Figure \ref{fig-outflow} 
    with crosses (``$\times$"). The spatial distribution of the 
    44 GHz masers shows four distinct groups. 
    The southern group of masers are distributed mainly along a N-S line 
    closely associated with the R1 lobe. The 
    northeast group of masers is spatially coincident with the B2 lobe and 
    the extended 4.5 \micron\ emission. In the north, a crowd of masers are 
    located in between the B1 and B2 lobes. There are also several maser spots 
    on the edge of the B1 lobe. A small group of masers detected in the southwest 
    are not consistent with any high-velocity CO emission. Noticeably, the 
    velocities of maser spots generally agree with 
    those of the associated outflow lobes, supporting their shock-driven 
    origin. There is a reversal of maser velocities at the southern tip 
    of the R1 lobe coincident with a change of outflow direction from
    southwards to the southeast. This 
    intriguing feature could be due to the influence of the adjacent 
    infrared bubble (see Section \ref{sec-discussion-bubble}).  

	Physical parameters of the outflows have been calculated from 
	the CO line wings under LTE conditions and assuming optically thin emission.
	The CO column density of each pixel in each velocity channel 
	can be straightforwardly
	derived from Equation \ref{eq-lte-density} as:
	\begin{eqnarray}
		\mathrm{d}N_\mathrm{CO}\mathrm{(cm^{-2})} &=&
		2.49\times10^{14}(T_\mathrm{ex}+0.92)\exp(\frac{17}{T_\mathrm{ex}}) \nonumber\\
		& &\times\frac{J(T_\mathrm{ex})}{J(T_\mathrm{ex})-0.018}T_\mathrm{r}~\mathrm{d}v \nonumber\\
		&=&F(T_\mathrm{ex})T_\mathrm{r}~\mathrm{d}v,
	\end{eqnarray}
	where $ \mathrm{d}v $ is the velocity interval in \kms and $ T_\mathrm{r} $ is 
	the measured brightness temperature in K. In the calculation, the excitation 
	temperature has been assumed to be approximately 
	the source-averaged dust temperature of clump C1 which is 21 K.
	The outflow mass, momentum, energy, dynamical age, mass loss rate,
	and mechanical force have been obtained from:
	\begin{equation}
		M = \mu_\mathrm{H_2}m_\mathrm{H}F(T_\mathrm{ex})\frac{d^2}{X_\mathrm{CO}}
		    \int_{\Omega}\int_{v}T_\mathrm{r}~\mathrm{d}\Omega\mathrm{d}v,
	\end{equation}
	\begin{equation}
		P = \mu_\mathrm{H_2}m_\mathrm{H}F(T_\mathrm{ex})\frac{d^2}{X_\mathrm{CO}}
			\int_{\Omega}\int_{v}T_\mathrm{r}v~\mathrm{d}\Omega\mathrm{d}v,
	\end{equation}
	\begin{equation}
		E = \frac{1}{2}\mu_\mathrm{H_2}m_\mathrm{H}F(T_\mathrm{ex})\frac{d^2}{X_\mathrm{CO}}
			\int_{\Omega}\int_{v}T_\mathrm{r}v^2~\mathrm{d}\Omega\mathrm{d}v,
	\end{equation}
	\begin{equation}
		t_\mathrm{dyn} = \frac{L_\mathrm{flow}}{v_\mathrm{char}},
	\end{equation}
	\begin{equation}
		\dot{M}_\mathrm{out} = \frac{M}{t_\mathrm{dyn}},
	\end{equation}
	\begin{equation}
		F_\mathrm{mech} = \frac{P}{t_\mathrm{dyn}}.
	\end{equation}
	Here, $\Omega$ is the total solid angle that the flow subtends, $v$ is
	the flow velocity with respect to the systemic velocity, 
	${v_\mathrm{char}} = P/M$
	is the characteristic outflow velocity, and $ L_\mathrm{flow} $ is the flow 
	length.
	
	The resultant outflow parameters are given in Table \ref{tb-outflow}. 
    The four lobes have dynamical ages ranging from $0.47\times10^4$ 
    to $1.33\times10^4$ yr, comparable to those of other outflows in
    early high-mass
    star-forming regions, e.g., 
    G240.31+0.07 \citep[$2.4\times10^4$ yr,][]{2009ApJ...696...66Q}, 
    G24.78+0.08 
    \citep[$\sim2\times10^4$ yr,][]{2011A&A...532A..91B}, 
    G28.34+0.06 \citep[$ (1.3-3.4)\times10^4 $ yr,][]{2011ApJ...735...64W}, 
    G11.11-0.12 \citep[$ \sim2.5\times10^4 $ yr,][]{2014MNRAS.439.3275W}, and
    G9.62+0.19 MM6 \citep[$ 8.5\times10^{3} $ yr,][]{2017arXiv170504907L}.
    The relatively small dynamical ages are also consistent 
    with the stellar age estimate from SED fitting ($ [0.6-3.8]\times10^4 $ yr,
    see Section \ref{sec-yso-sed}). 


    \subsection{Millimeter Methanol Maser Emission} \label{sec-results-maser}

    \begin{figure*}[htb]
	\centering
	\includegraphics[width=0.95\textwidth]{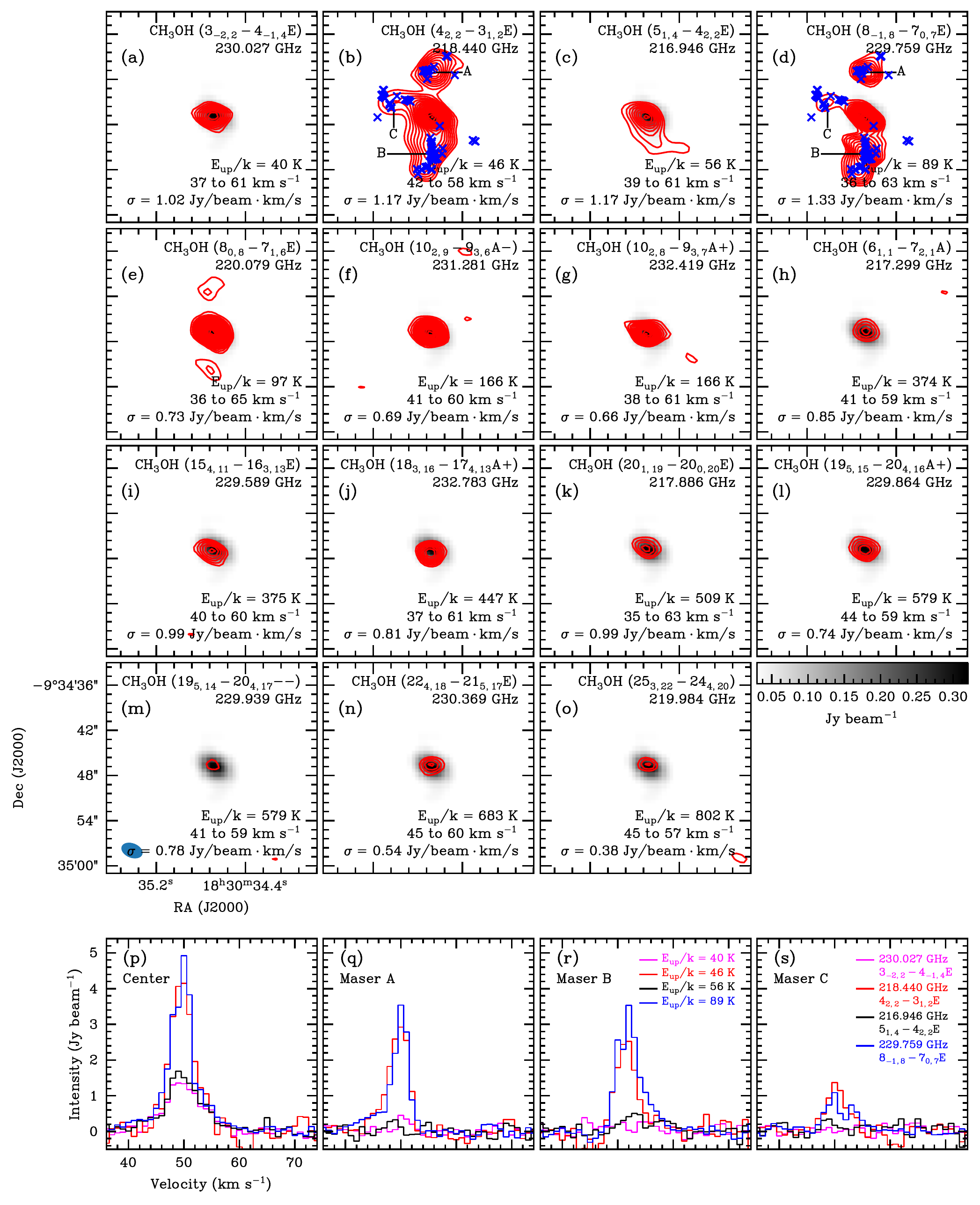}
	\caption{{\it(a)-(o)}: Velocity integrated intensities (contours)  of
		all the 15 \methanol~transitions 
		with lowest upper energies overlaid on the 1.3 mm continuum image. 
		Contours for each line start from $3\sigma$ 
		with an increment of $1\sigma$. The relevant transitions, 
		velocity intervals for integration, and $1\sigma$ noise levels
		are labeled at the bottom individual panels. Panels {\it(p)-(s)} presents spectra of \methanol~($3_{-2,2}-4_{-1,4}$), \methanol~($4_{2,2}-3_{1,2}$), \methanol~($5_{1,4}-4_{2,2}$), and \methanol~($8_{-1,8}-7_{0,7}$) at core SMA1, and positions A, B and C labeled in \it{(b)} and {\it(d)}. \label{fig:methanol}}
	\end{figure*}

    {Figures \ref{fig:methanol} {\it(a)-(o)} show the velocity integrated 
    intensity maps of all the 15 detected transitions of 
    \methanol~with upper energies ranging 
    from 40 K to 802 K. }
	All \methanol~transitions, with the exception of the 
    $4_{2,2}-3_{1,2}\ E$ (at 218.440 GHz with $E_\mathrm{up}/k=46$ K) 
    and $8_{-1,8}-7_{0,7}\ E$ (at 229.759 GHz with $E_\mathrm{up}/k=89$ K), 
    show {single core} morphologies. Emission from the 
    $4_{2,2}-3_{1,2}\ E$  
    and $8_{-1,8}-7_{0,7}\ E$  
    transitions reveals more complex structures. Figure
    \ref{fig:methanol} shows maps of the four \methanol~transitions with 
    lowest upper energies. Compared to the other methanol 
    transitions, strong 
    \methanol~($8_{-1,8}-7_{0,7}$) and \methanol~($4_{2,2}-3_{1,2}$) 
    emission can be detected to the north, south, and northeast of 
    the 1.3 mm core SMA1, spatially coincident with Class I 44 GHz methanol masers, 
    suggestive of maser
    emission in the \methanol~($8_{-1,8}-7_{0,7}$) and 
    \methanol~($4_{2,2}-3_{1,2}$) transitions.

    \methanol~($8_{-1,8}-7_{0,7}$) and 
    \methanol~($4_{2,2}-3_{1,2}$) spectra towards core SMA1, and positions A, B and C 
    marked in Figures \ref{fig:methanol} {\it(b)} and {\it(d)} 
    are presented in Figures \ref{fig:methanol} {\it(p)}-{\it(s)}. 
    For comparison, spectra of \methanol~($3_{-2,2}-4_{-1,4}$) 
    at 230.027 GHz with $E_\mathrm{up}/k=40$ K and \methanol~($5_{1,4}-4_{2,2}$) 
    at 216.946 GHz with $E_\mathrm{up}/k=56$ K are also shown.  
    \methanol~($8_{-1,8}-7_{0,7}$) and \methanol~($4_{2,2}-3_{1,2}$) spectra at 
    position A are narrow, resembling features of maser emission. 
    Relatively broader spectra at positions B and C may be 
    due to the blending of unresolved maser spots.

    The peak intensities of \methanol~($8_{-1,8}-7_{0,7}$) at positions SMA1,
    A, B, and C are 4.9, 3.5, 3.5, and 1.1 Jy beam$^{-1}$. For \methanol~($4_{2,2}-3_{1,2}$), 
    the peak intensities at SMA1, A, B, and C are 4.1, 2.9, 2.5, and 1.4 Jy beam$^{-1}$.
    The peak intensities of \methanol~($3_{-2,2}-4_{-1,4}$) at SMA1, A, B, and C
    are 1.4, 0.4, 0.2, and 0.1 Jy beam$^{-1}$.
    The $8_{-1,8}-7_{0,7}/3_{-2,2}-4_{-1,4}$ line ratio has 
    previously been used as a discriminant between 
    thermal and non-thermal $8_{-1,8}-7_{0,7}$ emission, with ratios $>3$ 
    indicative of non-thermal 
    \citep{2002ARep...46...49S,2011ApJ...729..124C}. The $8_{-1,8}-7_{0,7}/3_{-2,2}-4_{-1,4}$ 
    ratios for SMA1, A, B, and C are 3.5, 8.8, 17.5, and 11.0, suggestive of 
    non-thermal $8_{-1,8}-7_{0,7}$ emission at positions A, B, and C. 
    We can similarly use the
    $4_{2,2}-3_{1,2}/3_{-2,2}-4_{-1,4}$ line ratio to distinguish thermal from 
    non-thermal $ 4_{2,2}-3_{1,2} $ 
    emission. Under LTE and optically thin conditions, the thermal 
    emission ratio of two \methanol~transitions can be obtained from
    \begin{equation}
    R\left(\frac{\nu_1}{\nu_2}\right)=\left(\frac{A_\mathrm{ul1}}{A_\mathrm{ul2}}\right)
    \left(\frac{\nu_2}{\nu_1}\right)^2\exp
    \left(\frac{E_\mathrm{up2}-E_\mathrm{up1}}{k T}\right).
    \end{equation}
    With an assumed temperature of 21 K (similar to the dust temperature of
    clump C1) deviating from the 1.3 mm core, the 
    thermal emission $4_{2,2}-3_{1,2}/3_{-2,2}-4_{-1,4}$ ratio will
    be approximately 3. Thus, we suggest a ratio threshold of 3 as a demarcation 
    of thermal and non-thermal 218.440  GHz emission. The 
    $4_{2,2}-3_{1,2}/3_{-2,2}-4_{-1,4}$ 
    ratios for SMA1, A, B, and C are 2.9, 7.3, 12.5, and 14.0, supportive of 
    non-thermal 218.440 GHz emission at positions A, B, and C. 
    The \methanol~($8_{-1,8}-7_{0,7}$) 
    and \methanol~($4_{2,2}-3_{1,2}$) emission features at positions A, B, 
    and C spatially coincide well with 44 GHz Class I \methanol~masers. 
    This agreement, along with large $8_{-1,8}-7_{0,7}/3_{-2,2}-4_{-1,4}$ and 
    $4_{2,2}-3_{1,2}/3_{-2,2}-4_{-1,4}$ 
    ratios and narrow line widths (especially for position A), gives strong 
    support to the interpretation of millimeter \methanol~maser emission.

    Maser emission of \methanol~($8_{-1,8}-7_{0,7}$) at 229.759 GHz has been 
    discovered by \citet{2002ARep...46...49S} in DR21(OH) and DR21 West, 
    and reported in HH 80-81 \citep{2009ApJ...702L..66Q}, IRAS 05345+3157 
    \citep{2009A&A...499..233F}, NGC 7538 \citep{2011ApJ...728....6Q}, and 
    EGOs G11.92-0.61, G18.67+0.03 and G19.01-0.03 
    \citep{2011ApJ...729..124C,2012ApJ...760L..20C}. However, maser emission 
    of the \methanol~($4_{2,2}-3_{1,2}$) transition at 
    218.440 GHz has, to our knowledge, only been tentatively detected in DR21(OH) by \citet{2012ApJ...744...86Z}
    and in NGC 6334 I(N) by \citet{2014ApJ...788..187H}
    prior to this work. Population inversion for the 218.440 GHz 
    transition has been predicted in maser models \citep{2012IAUS..287..433V}. 
    The $J_2-(J-1)_1$ transition and association with 44 GHz class I masers 
    suggest that the 218.440 GHz \methanol~masers detected in G22 are Class 
    I type. The small number of known sources exhibiting maser emission 
    in this transition is likely to be because few previous observations 
    of high-mass star formation regions have covered this line.
    
\section{Discussion}\label{sec-discussions}

\subsection{Global Collapse of the Cloud}

	\begin{figure}[htb]
		\centering
		\includegraphics[width=0.45\textwidth]{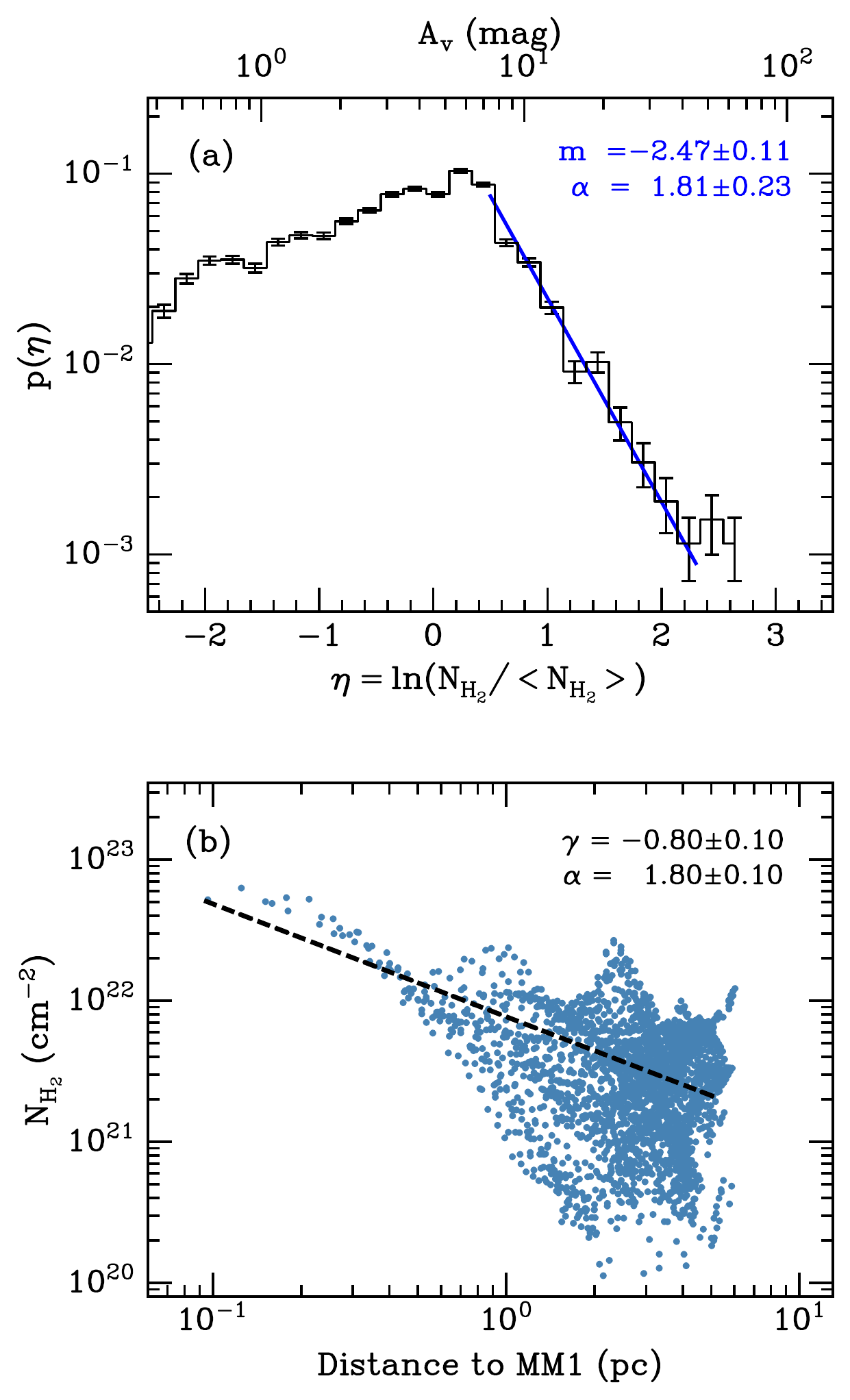}
		\caption{\textit{(a)} Probability distribution function (PDF) 
			of $ N_\mathrm{H_2} $. {The 
			blue solid line shows a power-law fit
			to the high-column density part of
			the PDF. Here, $<N_\mathrm{H_2}>=4.2\times10^{21}$ cm$^{-2}$.
			A conversion factor of 
			$ N_\mathrm{H_2}/A_\mathrm{v}=9.4\times10^{20} $ cm$^{-2}$ mag$^{-1}$ 
			is used to convert column density to 
			visual extinction.}
			\textit{(b)} 
			Radial column density profile. Each point represents one 
			pixel value in the map. The \textit{x}-axis gives the
			distance to clump C1. The dashed line shows a power-law
			fit. The exponent $ \gamma $ and the exponent ($ \alpha $) of 
			the underlying radial number density profile
			$ \rho\propto r^{-\alpha}\propto r^{\gamma-1} $
			are also labeled. \label{fig-Nh2-pdf}}
	\end{figure}

	\subsubsection{Density Structure}
	
	The probability distribution function (PDF) of $N_\mathrm{H_2}$ 
	is shown in Figure \ref{fig-Nh2-pdf} (a). {The high-density end of
	 the $N_\mathrm{H_2}$-PDF 
	of G22 can be fitted to a power-law.} Such $N_\mathrm{H_2}$-PDF profiles 
	have been reported in many IRDCs and IR-bright clouds 
	\citep[e.g.,][ and references therein ]{2015A&A...578A..29S}.
	{Power-law 
	tails in $N_\mathrm{H_2}$-PDFs} also have been observed in numerical works
	\citep[e.g.,][]{2008ApJ...688L..79F,2010A&A...520A..17K,2011ApJ...727L..20K}.
	In both theoretical and observational studies, self-gravity has
	been suggested to be the dominant process in the formation of the power-law
	tail. Assuming a spherical symmetry, the power-law slope $ m $ of the $N_\mathrm{H_2}$-PDF
	is related to the exponent $ \alpha $ of a radial density profile of 
	$ \rho(r)\propto r^{-\alpha} $ and $ \alpha=-2/m+1 $ 
	\citep{2013ApJ...763...51F}. An $ \alpha $ of 
	about {$ 1.81\pm0.23 $} is consistent with
	self-gravity \citep{2014ApJ...781...91G,2015A&A...575A..79S}. 
	Such self-gravity can be attributed to local free-fall of individual 
	clumps/cores or global collapse.
	
	Figure \ref{fig-Nh2-pdf} (b) shows the radial column density profile
	which can be well fitted with a power-law 
	$ N\propto\rho(r)\times r\propto r^{1-\alpha}\propto r^\gamma$.
	Here $ \alpha $ is the exponent of the underlying radial number density profile.
	The resultant {$ \alpha=1.80\pm0.10 $} is comparable to the $ \alpha $ from the power-law
	tail of the $N_\mathrm{H_2}$-PDF, consistent with gravitational collapse on large
	scales. This scenario is further supported by large scale velocity gradients discussed
	in the following section.

	\begin{figure}
		\centering
		\includegraphics[width=0.48\textwidth]{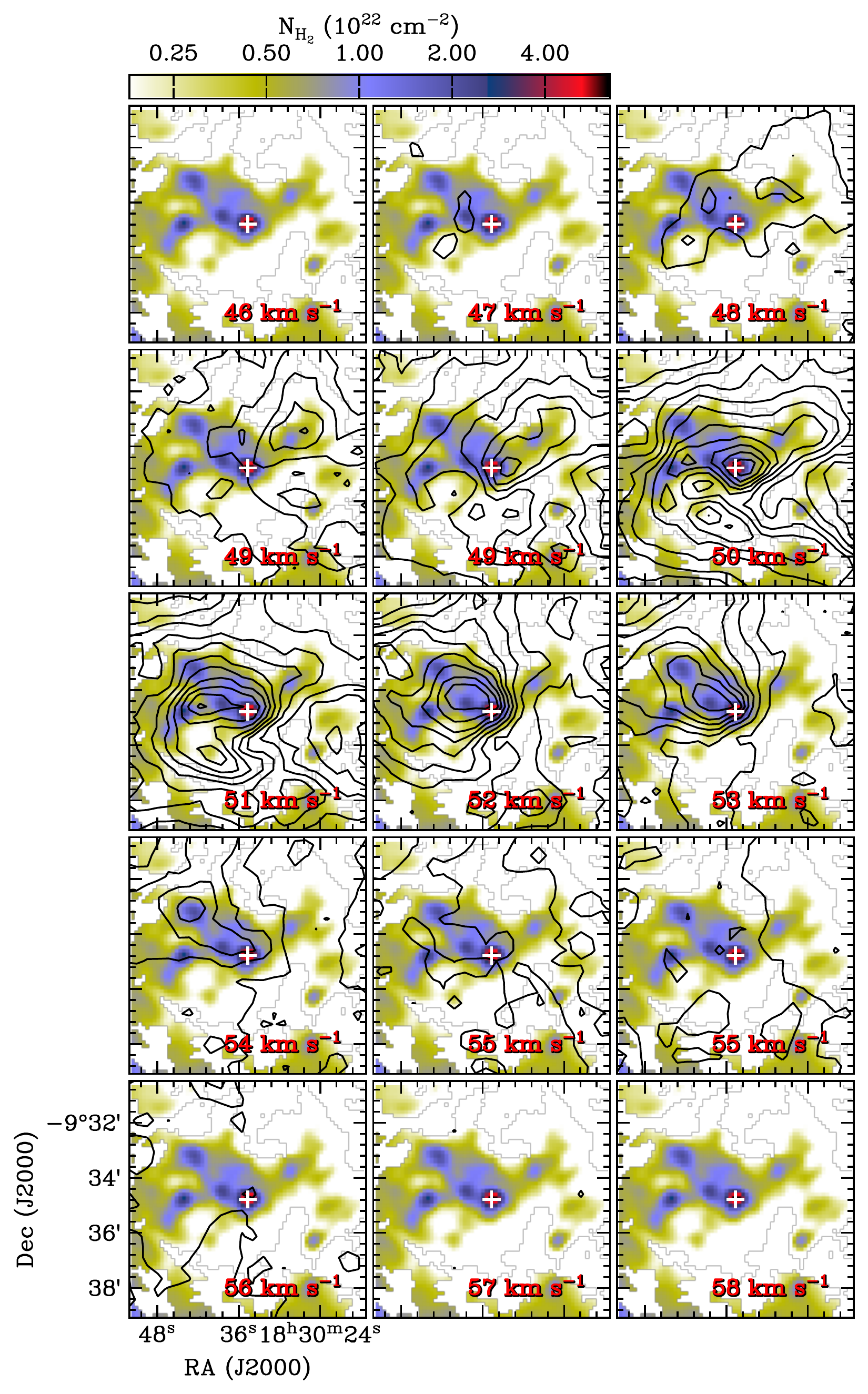}
		\caption{GRS \tco~$ (1-0) $ channel maps overlaid on
			the column density map from SED fits. 
			The velocity interval of integration for each subplot 
			covers 1 \kms~with the center value marked in the 
			lower-right corner. \tco~Contours start 
			from $4\sigma$ and increase with
			a step of $ 7\sigma $. Here, $ 1\sigma=0.05 $ K \kms. \label{fig-GRS-Channel}}
	\end{figure}

	\begin{figure*}[htb]
		\centering
		\includegraphics[width=0.95\textwidth]{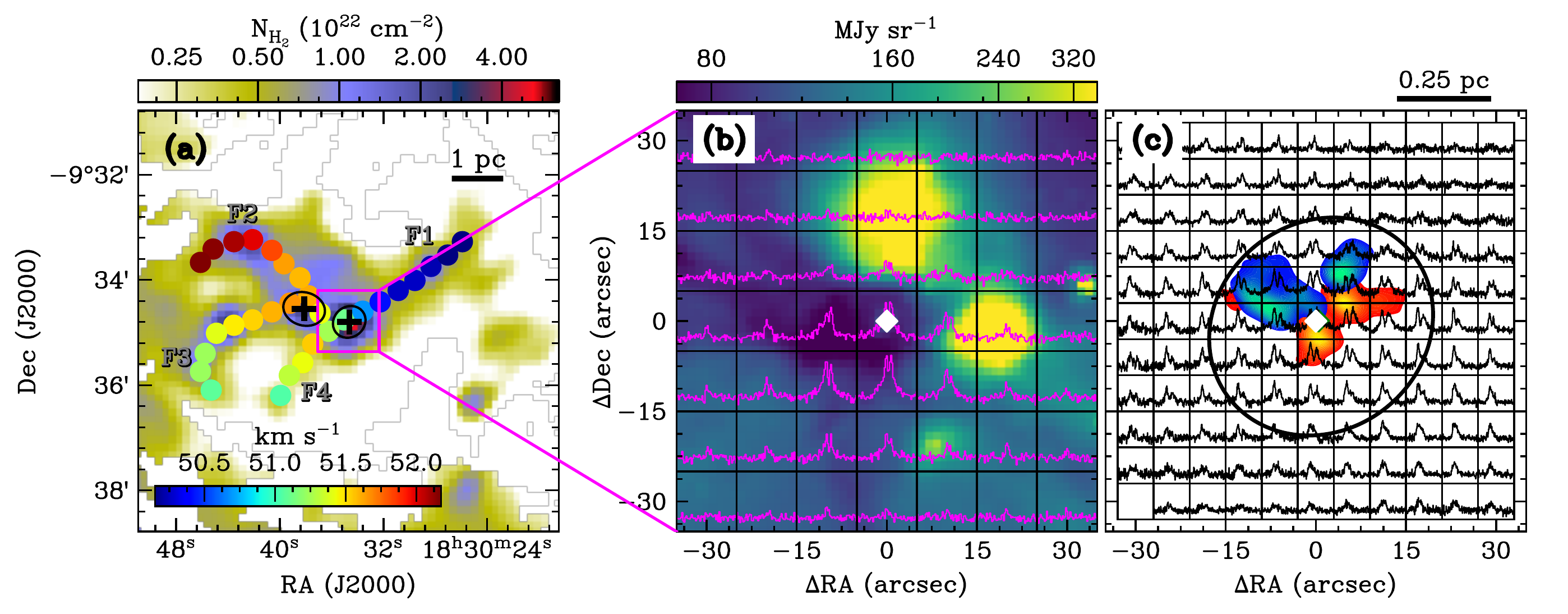}
		\caption{\textit{(a)} Velocity centroids of \tco~$(1-0) $ extracted along filaments
			overlaid on top of a $ N_\mathrm{H_2} $ column density map. 
			Also labeled are the two most massive clumps (black ``+'' and small ellipses) 
			in the hub region.
			\textit{(b)} Spectra of JCMT/\hcop~$ (3-2) $ overlaid on the IRAC 8.0 
			\micron~image.
			\textit{(c)} Spectra of JCMT/\tco~$ (3-2) $ overlaid on the 
			SMA/CO $ (2-1) $ outflows. The white diamond in panels (b) and
			(c) labels the 1.3 mm core SMA1. The large ellipse in (c) 
			delineates clump C1.
			\label{fig-infall-image}}
	\end{figure*}

	\begin{figure}[htb]
	\centering
	\includegraphics[width=0.45\textwidth]{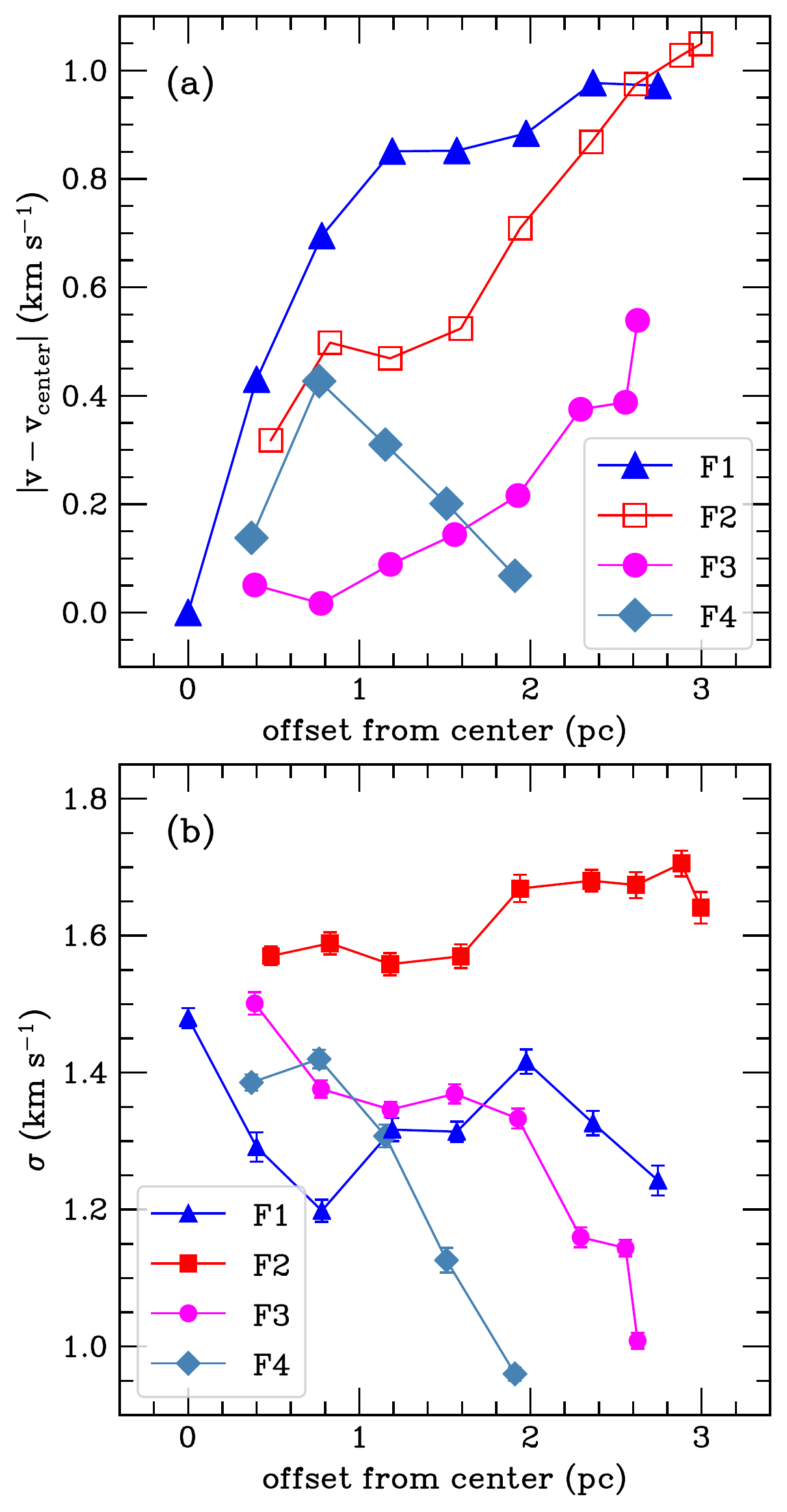}
	\caption{(a) Line-of-sight velocity of \tco~$ (1-0) $ as a function
		of position from the potential well centers, i.e., clump C1 for F1, F2, and F4, 
		and clump C2 for F3. {(b) Velocity dispersion of \tco~$ (1-0) $
		as a function of offset from the potential well centers. }\label{fig-vel-prof}}
	\end{figure}

	\subsubsection{Filamentary Collapse}
	
	Assuming cylindrical hydrostatic equilibrium, a critical mass per unit length 
	$ M_\mathrm{line} $
	has been proposed to be $ M_\mathrm{line,crit}=2c_\mathrm{s}^2/G
	=16.7\left(\frac{T}{10~\mathrm{K}}\right)~M_\odot~\mathrm{pc}^{-2}$ 
	\citep{1964ApJ...140.1056O,1997ApJ...480..681I,2013ApJ...766..115K,2014prpl.conf...27A}.
    With a typical temperature of about 20 K (see Section \ref{sect-largeDust} and 
    Figure \ref{fig-Nh2-Tdust}), the critical $ M_\mathrm{line,crit} $ value is about 33 
    \msun~pc$^{-1}$ for the filaments in G22. All four filaments are supercritical 
    with $ M_\mathrm{line} $ 
    significantly larger than the critical value, consistent with the velocity measurements
    showing that gas is globally infalling in these filaments (see bellow). 
    The mass per unit length for the four filaments ranges from {54 to 220} 
    \msun~pc$^{-1}$ (see Table \ref{tb-filaments}). The $ M_\mathrm{line} $ of filaments in G22 are comparable 
    to those in DR21 where gas is channeled along the filaments onto the central
    ridge region \citep{2012A&A...543L...3H}.
    
    GRS \tco~$ (1-0) $ channel maps are shown in Figure \ref{fig-GRS-Channel}.
    Intriguingly, emission spatially associated with filament F1 mainly originates from
    gas with velocity smaller than 51 \kms~while the gas associated with filaments F2 and F3 
    is systematically red-skewed with velocities larger than 50 \kms. The variation of
    velocity indicates bulk motions in these filaments.
    
    Spectra of \tco~$ (1-0) $ have been extracted along the filaments 
    with a step-size of half the angular resolution ($ \sim46\arcsec $) 
    of the GRS observations,
    and fitted to a one-dimensional Gaussian profile
    using the \textit{GILDAS/CLASS} software. The resultant centroid velocities of the
    extracted spectra are shown in Figure \ref{fig-infall-image} (a) as color-coded
    circles. The smoothly varying velocity along filaments F1, F2, and F3 is 
    reminiscent of the situation in the SDC13 infrared dark clouds where 
    longitudinal filamentary collapse has been reported \citep{2014A&A...561A..83P}.
    
    Alternative possible physical processes, which can lead to
    the observed velocity pattern, include collapse,  rotation,
    filament collision, expansion, and wind-driven acceleration \citep{2014A&A...561A..83P}.
    The observed velocity gradients along the filaments in G22 do not allow
    us to determine a common axis for rotation to take place unless the axis goes
    through the filament junction. However, such a scenario is unrealistic as
    differential rotation would tear apart these filaments \citep{2014A&A...561A..83P}.
    No signs of enhanced linewidth in the interaction and bridging features 
    \citep{2003PASP..115..953B,2017ApJ...835L..14G} can 
    be observed in the junction, 
    excluding the possibility of filament collision. 
    No \hii~region is observed at the center of G22. And the outflows 
    driven by MIR1 in clump C1 cannot explain the red-skewed velocity along
    F2 as the outflowing gas in that direction is blue-shifted (see Figure \ref{fig-outflow}).
    Similar to the situation in SDC13 \citep{2014A&A...561A..83P}, 
    there also resides a mid-IR bright nebula to the southwest of G22 
    (see Figures \ref{fig-mul-morph}). This mid-IR nebula corresponds to an 
    infrared bubble MWP1G022027+002159 and is interacting with G22 (see
    Section \ref{sec-discussion-bubble}). As MWP1G022027+002159 is located
    slightly behind G22 (see Section \ref{sec-discussion-bubble}), 
    its expansion would lead to systematic 
    blue-shifted velocities in the interaction region, but the affected area is limited.
    The largely blue-skewed velocity in the most distant section of F1 and 
    the red-skewed velocity
    in F2 cannot be explained by wind-driven expansion. 
    
    Figure \ref{fig-vel-prof} (a) shows the difference between the 
    velocities of the filaments and the junction as a function 
    of distance to the center. Monotonically increasing profiles
    for F1, F2, and F3 are consistent with those observed in other reported
    collapsing filaments
    \citep[e.g., SDC13 and AFGL 5142 in][]{2014A&A...561A..83P,
    	2016ApJ...824...31L}. 
    The estimated velocity gradients in F1, F2 and F3 are 
    0.36, 0.35 and 0.21 \kms~pc$^{-1}$, which are comparable to
    those in SDC13 \citep[$ 0.22-0.63 $,][]{2014A&A...561A..83P}
    , and smaller than those in The Serpens South cluster 
    \citep[$ \approx1.4 $,][]{2013ApJ...766..115K} and  
    AFGL 5142 \citep[$ 9-17 $,][]{2016ApJ...824...31L}.
    The small velocity gradients in G22 may indicate low
    filament inclination.
    
    {The velocity dispersion of the filaments as a function
    of distance to the center is shown in Figure  \ref{fig-vel-prof} (b).
    For filaments F3 and F4, an increasing trend of velocity dispersion 
    can be seen toward the center region. This increase could be 
    due to additional kinetic energy converted from gravitational energy
    during the collapse of the filaments \citep{2014A&A...561A..83P}. Such
    increase of velocity dispersion can also be observed toward the inner part 
	of filament F1. In contrast, filament F2 has significantly larger velocity
	dispersions. As \tco~$ (1-0) $ could be optically thick, observations 
	of more optically thin lines, such as N$_2$H$^+$, would help better reveal
	the velocity dispersions.}

    Following \citet{2013ApJ...766..115K}, we estimated the
    mass infall rates $ \dot{M} = \frac{\bigtriangledown V M}{\tan(\alpha)}$
    to be {132, 206 and 104 \msun~Myr$^{-1}$ in} F1, F2, and F3.
    Here, we assumed an inclination angle of 45$\deg$. Therefore,
    about {440} \msun~gas will be channeled to the junction regions
    in $ 1 $ Myr if the current accretion rates are sustained. 
    This high accretion will double the mass of the hub region
    in about 6 free-fall times where the free-fall time is about 
    0.31 Myr. 
    Although two million years of high-accretion may not be plausible, 
    according to models of the evolution of high-mass star, 
    it is tenable to posit that a fraction of
    the masses in the central
    clumps C1 and and C2 have been assembled through large-scale filamentary 
    collapse.

    \subsection{Clump-Fed Accretion in Clump C1}\label{sec-discuss-clump-fed}
    
        \begin{figure}
    	\centering
    	\includegraphics[width=0.48\textwidth]{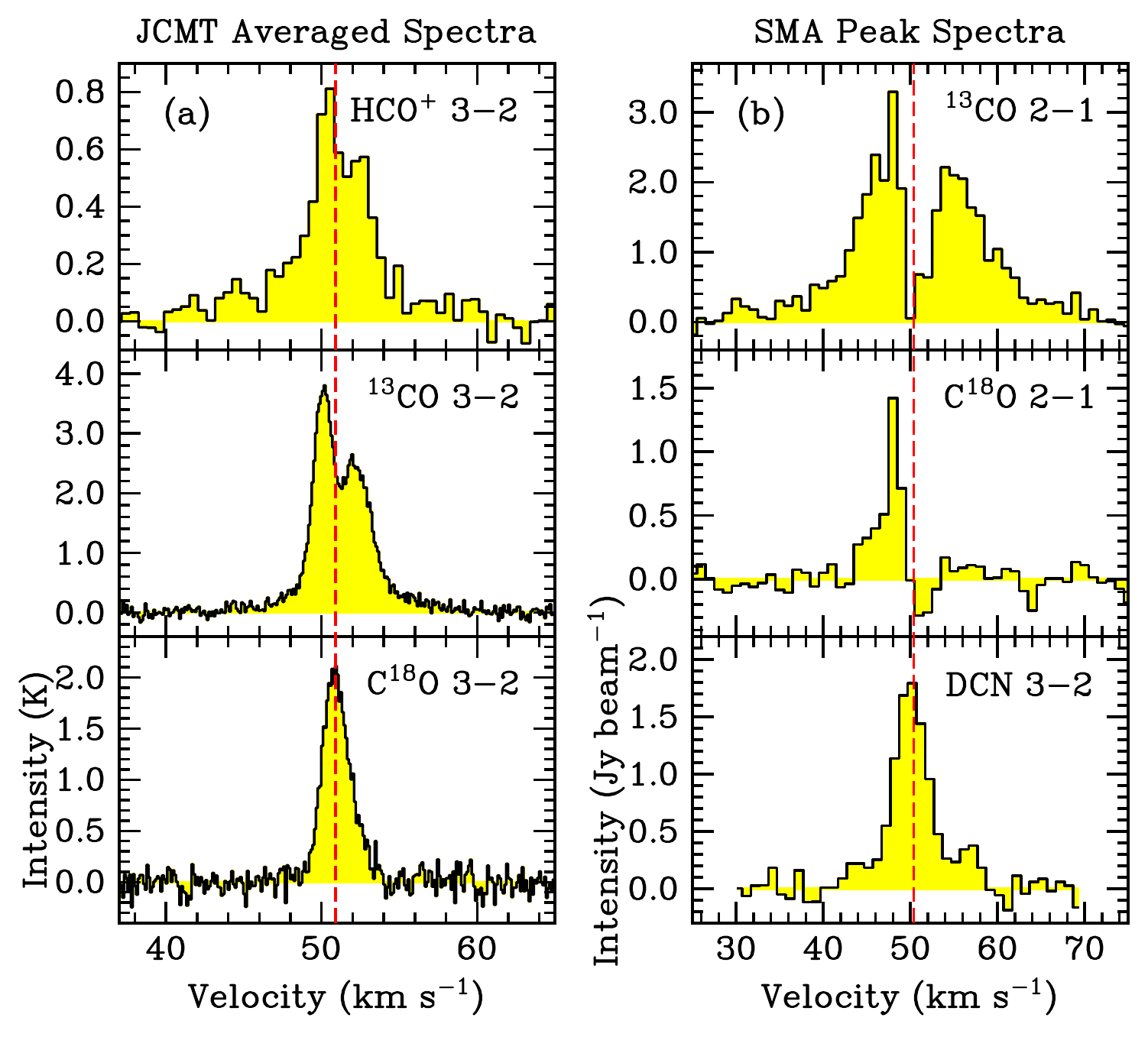}
    	\caption{\textit{(a)} JCMT \hcop~~$ (3-2) $, \tco~$ (3-2) $, and \ceo~$ (3-2) $
    		spectra 
    		averaged over clump C1.  \textit{(b)} SMA \tco~$ (2-1) $, and \ceo~$ (2-1) $
    		and  DCN~$ (3-2) $ spectra at the peak of SMA1.
       		\label{fig-infall-tracers}}
    \end{figure}
%
%

	\begin{figure*}[htb]
		\centering
		\includegraphics[width=0.95\textwidth]{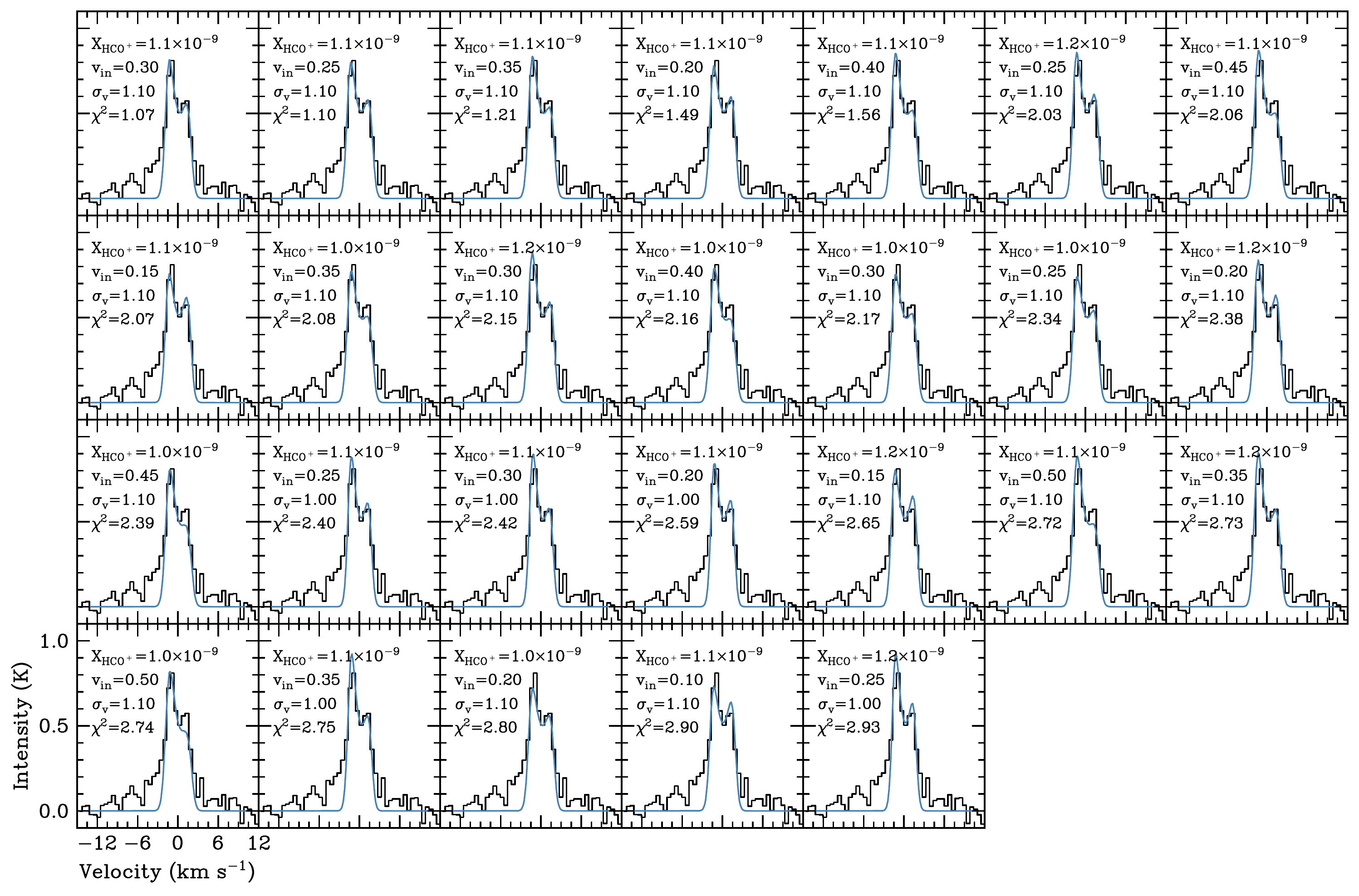}
		\caption{Spectrum of \hcop~$ (3-2) $ averaged over clump C1. 
			The \hcop~$ (3-2) $ spectra obtained from 
			RATRAN modeling of a collapsing clump are shown
			as blue. Details about the modeling are given 
			in Section \ref{sec-discuss-clump-fed}. Here, we only show the models with reduced
			$ \chi^2 $ smaller than 3. The \hcop~abundance, infall
			velocity, and velocity dispersion are labeled for each 
			model.
			\label{fig-infall-model}}
	\end{figure*}

    A virial parameter of $ \alpha_\mathrm{vir} = 5\sigma^2_\mathrm{turb}R/GM = 0.58$ 
    for clump C1
    has been estimated based on a velocity dispersion of $ \sigma_\mathrm{turb} = 0.87$
    \kms~measured from the JCMT \ceo~$ (3-2) $ data. With an $ \alpha_\mathrm{vir} $ smaller 
    than 1, C1 is likely dominated by gravity and potentially collapsing. Magnetic fields might
    provide additional support against gravity. Following \citet{2011A&A...530A.118P},
    we estimated that the magnetic field strength necessary to virialize C1 is 
    $ B_\mathrm{vir} = 260$ $ \mu$G, which is comparable to estimates 
    for the field in clouds at similar densities 
    \citep{2012ARA&A..50...29C,2014prpl.conf..101L}. 
    Note that the virial parameter
    may have been overestimated, as any systematic motions, such as infall and outflow,
    could significantly increase the velocity-dispersion estimate.
    
    Figure \ref{fig-infall-image} (b) and (c) show JCMT \hcop~$(3-2) $
    and \tco~$ (3-2) $ spectra of clump C1 overlaid on the IRAC/8.0 \micron~image and 
    SMA CO $ (2-1) $
    outflows. All spectra with significant emission ($ >3\sigma $) for both \hcop~$ (3-2) $
    and \tco~$ (3-2) $ show blue-skewed self-absorbed profiles, with the exception of
    some spectra 
    at the center (for both lines) and the north (for \tco~$(3-2) $). Such blue profiles
    could originate from collapsing or rotating clouds with an inward increasing temperature 
    profile. The different line profiles in the center could be due to the inclusion
    of emission from the outflow
    from SMA1. Although we cannot exclude the contribution of rotation, inward motions 
    must occur to explain the dominance of 
    blue profiles in both lines, especially in the clump-confined region (see the large
    circle in Figure \ref{fig-infall-image} (c)). This is further supported by 
    the blue-skewed clump-averaged spectra shown in Figure \ref{fig-infall-tracers} (a).
    {Intriguingly, the absorption dips of the source averaged \hcop~and 
    \tco~spectra are both red-skewed with respect to the systemic velocity. This feature
	is consistent with a global collapse scenario as suggested in 
	\citet{2016ApJS..225...21J}.}
    
    We estimated the infall velocity by fitting the clump-averaged  \hcop~spectrum
    using the RATRAN 1D Monte Carlo radiative transfer code \citep{2000A&A...362..697H}.
    The input parameters include the clump size, density profile, kinematic temperature
    profile, \hcop~fractional abundance ($ X_\mathrm{HCO^+} $)，and infall velocity.
    We deduced a power-law density profile ($ \rho\propto r^{-1.5} $) based 
    on the mass ($\sim590 $ \msun) 
    and radius ($ \sim0.39 $ pc) of C1. Following \citet{2013A&A...555A.112P},
    we used a constant temperature which is approximately the source-averaged 
    dust temperature of C1 (21 K, see Table \ref{tb-clumps}). {We ran a grid of
    880 models with varying \hcop~abundance $ X_\mathrm{HCO^+} $ in the range
    $ [0.6-1.5]\times10^{-9}$, 
    infall velocity $ v_\mathrm{in} $ in the range [$ 0.05-0.55 $]
    \kms, and velocity dispersion in the range $ [0.8-1.5] $ \kms}. 
	A reduced $ \chi^2 $ parameter
    was calculated for the averaged spectrum of each model. Note that only the 
    central 4 \kms~portion was considered in calculating reduced $ \chi^2 $ 
    as the emission from the outflows has a large impact on the outer channels of 
    the \hcop~spectrum. 
    {Fifty-six models with $ \chi^2-\chi^2_\mathrm{best}<3 $ 
    were considered to be
    good fits, here $ \chi^2_\mathrm{best} = 1.07 $}. Figure \ref{fig-infall-model}
    shows the averaged spectra of \hcop~$ (3-2) $ of models with $ \chi^2<3 $. 
    {From $ 1/\chi^2 $ weighted parameters of all models with good fits,
    we derived the infall velocity ($ V_\mathrm{in}$), 
    velocity dispersion ($ \sigma_\mathrm{v}$), and \hcop~abundance  $ X_\mathrm{HCO^+}$
    to be $ V_\mathrm{in} =0.31\pm0.12$ \kms, $ \sigma_\mathrm{v} = 1.10\pm0.05 $ \kms,
    and $ X_\mathrm{HCO^+} = (1.09\pm0.06)\times10^{-9}$, respectively.} We also investigated 
    the affect of varying the radius of the collapse $ R_\mathrm{in} $ and
    found that the \hcop~$ J=3-2 $ can be well modeled only for $ R_\mathrm{in}>0.5 $ pc.
    This is consistent with the observed blue-skewed \tco~$ J=3-2 $ spectra beyond the 
    extent of clump C1 (see Figure \ref{fig-infall-image} (c)) and 
    further supports a clump-scale collapse scenario.
    
    We estimated the current mass infall rate using 
    $ \dot{M}_\mathrm{in}  = 4\pi r^2  \rho v_\mathrm{in} = 1.5Mv_\mathrm{in}/r$
    to {be about $ 7.2\times10^{-4} $ \msun~yr$^{-1}$.} Here, $ \rho\propto r^{-1.5} $
    is assumed. {The mass infall rate is about five times smaller than the 
    free-fall accretion rate ($\dot{M}_\mathrm{ff} \sim3.55\times10^{-3} $ \msun~yr$^{-1}$)} 
	but still significantly larger than the mass outflow rate
    of SMA1, suggesting that the central hot core can keep growing
	in mass via clump-fed accretion. {Using
	a velocity dispersion of 0.87 \kms~measured from the JCMT \ceo~$(3-2)$,
	we estimate the crossing time of C1 to be 0.43 Myr which is
	about 2.5 free-fall times ($ t_\mathrm{ff} \sim 0.17 $ Myr). 
	In spite of a difference of about a factor 2, the crossing and free-fall
	times agree order-of-magnitude wise considering they are just approximate values. This 
	further supports a dynamical global collapse scenario for clump C1. }

    \subsection{SMA1: A Collapsing Hot Molecular Core} \label{sec-hot-core}
    
    \begin{figure}
    	\vspace{0.3cm}
    	\centering
    	\includegraphics[width=0.45\textwidth]{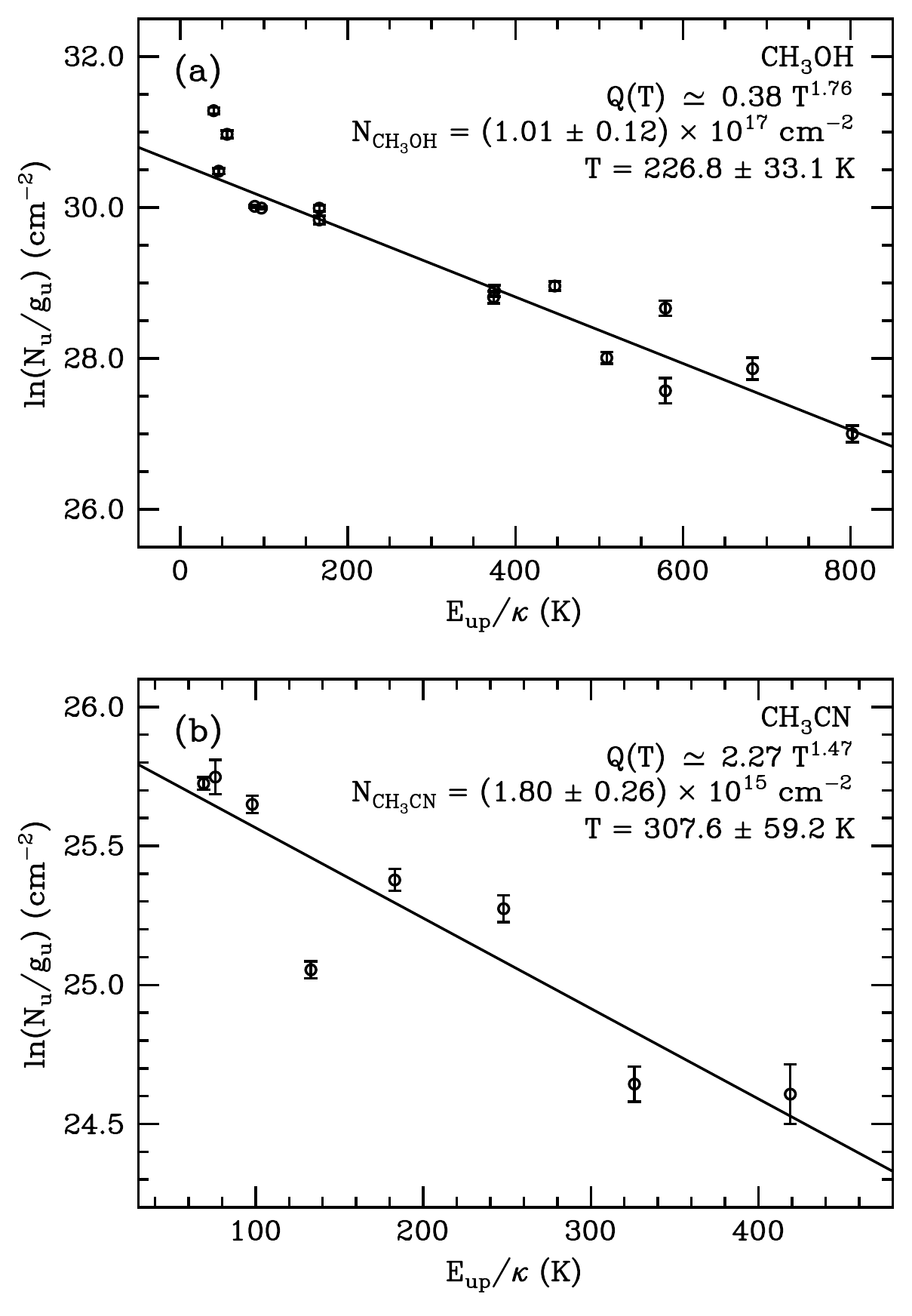}
    	\caption{Rotation diagrams of \methanol~\textit{(a)} and \ace~\textit{(b)}.\label{fig-rd}}
    \end{figure}
    
    Most of the 103 molecular lines detected towards the SMA1 peak
    are from complex organic molecules. This, together with the
    small size (0.034 pc) and high density ($ 7.2\times10^6 $ cm$^{-2}$), 
    suggests that SMA1 is
    a hot molecular core. The physical properties of hot cores
    are characterized by a small source size ($ \le0.1 $ pc),
    a high density ($ \ge10^{6} $ cm$^{-2}$), and warm gas/dust
    temperature ($ \ge100 $ K) 
    \citep{2000prpl.conf..299K,2004IAUS..221...59V}.

    \subsubsection{High Gas Temperature}\label{sec-temperature}

    Fifteen \methanol~transitions with $ E_\mathrm{up}/k $ ranging from 40 
    to 802 K and eight \ace~transitions with $ E_\mathrm{up}/k $ ranging from 
    69 to 419 K have been reliably detected toward SMA1. 
    Assuming that these lines are optically thin and SMA1
    is in LTE, beam-averaged temperatures 
    and column densities of \methanol~and \ace~can be obtained using the 
    rotational diagram method \citep{1999ApJ...517..209G}. The column density ($N_u$) 
    of the upper state of a transition can be expressed as a function of 
    its upper state energy($E_u/k$),
    \begin{equation}\label{eq-rd}
    \mathrm{ln}\frac{N_u}{g_u} = 
    \mathrm{ln}\frac{N}{Q(T_{\mathrm{rot}})}-\frac{E_u/k}{T_{\mathrm{rot}}},
    \end{equation}
    where
    \begin{equation}\label{eq-Nup}
    N_u = \frac{8\pi k\nu^2}{hc^3A_{ul}}\int T_bdv.
    \end{equation}
    Here, $g_u$ is the upper state degeneracy, $N$ is the beam-averaged total 
    column density of a species, $Q(T_{\mathrm{rot}})$ is the partition 
    function at a given temperature, $T_{\mathrm{rot}}$ is the rotation temperature, 
    $A_{ul}$ is the Einstein A-coefficient for the transition, and $\int T_bdv$ 
    is the velocity integrated intensity of a specific line.
    
    The partition function of \methanol~and \ace~can be expressed as,
    \begin{equation}
    Q(T_{\mathrm{rot}}) = aT_{\mathrm{rot}}^b.
    \end{equation}
    The constants of $a$  and $b$ were obtained by fitting 
    the JPL partition function values at 9.375 to 300 K. 
    
    With multiple transitions observed, which have upper state energies 
    spanning through a large range, the column density and rotation temperature 
    of a specific molecule can be obtained via a least-square fitting to 
    $\mathrm{ln}\frac{N_u}{g_u}$ as a linear function of $E_u/k$.
    
    The fitting results are shown in Figure \ref{fig-rd}. A rotation 
    temperature of $227\pm33$ K and a column density of $(1.01\pm0.12)\times10^{17}$ 
    cm$^{-2}$ have been attained for \methanol. The resulting temperature and 
    column density for \ace~are $308\pm59$ K and $(1.80\pm0.26)\times10^{15}$ cm$^{-2}$.
    The gas temperatures of SMA1 are consistent with those of 
    hot cores in our Galaxy \citep[120--480 K,][]{2014ApJ...786...38H}.

    \subsubsection{MIR1: an Embedded High-Mass Protostar}  \label{sec-yso-sed}
    
    \begin{figure}
    	\vspace{0.3cm}
    	\centering
    	\includegraphics[width=0.45\textwidth]{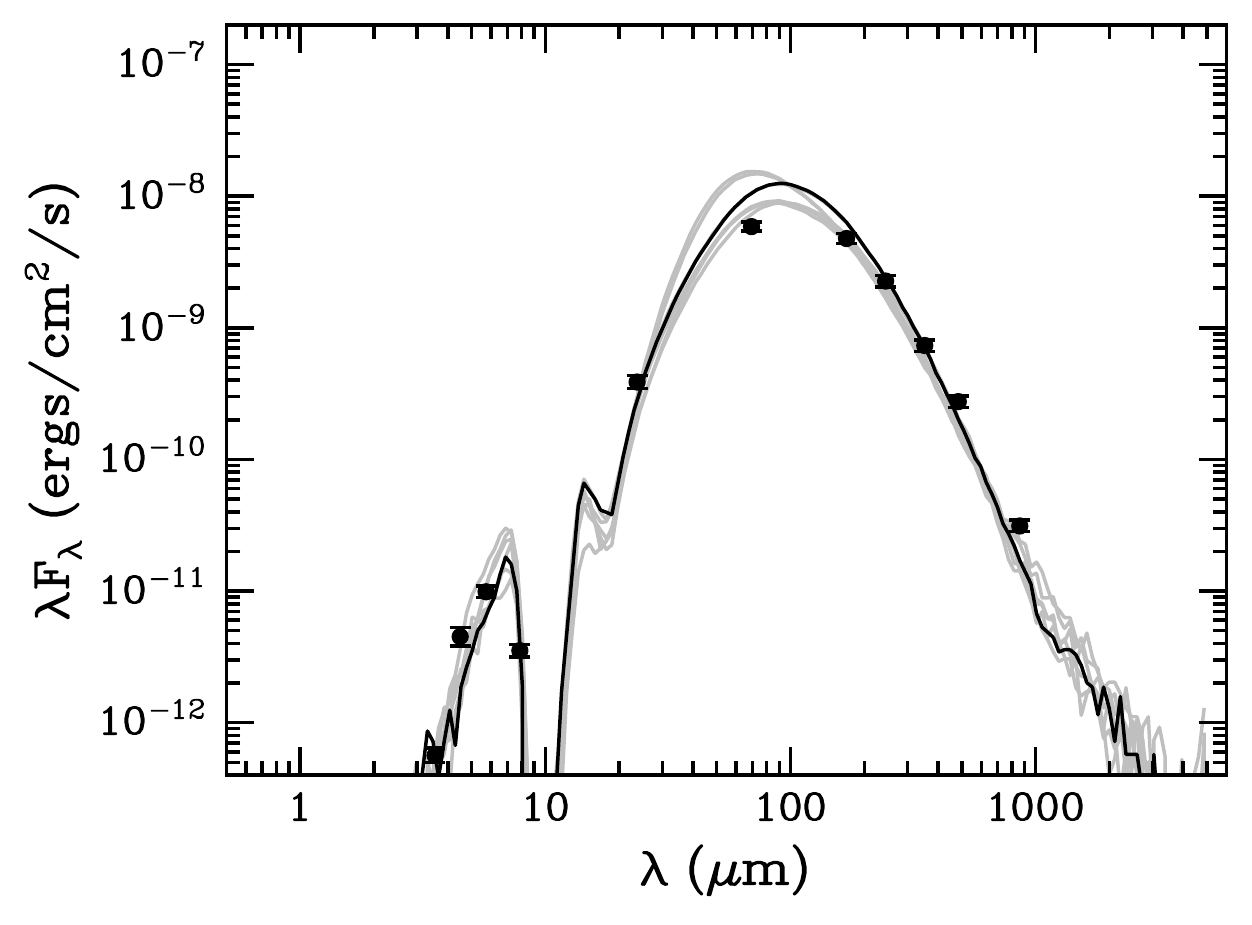}
    	\caption{Spectral energy distribution of MIR1 
    		(SSTGLMC G022.0387+00.2222). Photometric 
    		data from GLIMPSE, MIPSGAL, 
    		Hi-GAL, and ATLASGAL were fitted using YSO models
    		of \citet{2006ApJS..167..256R,2007ApJS..169..328R}.
    		The black line shows the best fit, and the gray lines 
    		show subsequent good fits with $ (\chi^2-\chi_{best}^{2})<3N $.
    		\label{fig-sed}}
    \end{figure}

	\begin{figure}
		\vspace{0.3cm}
		\centering
		\includegraphics[width=0.49\textwidth]{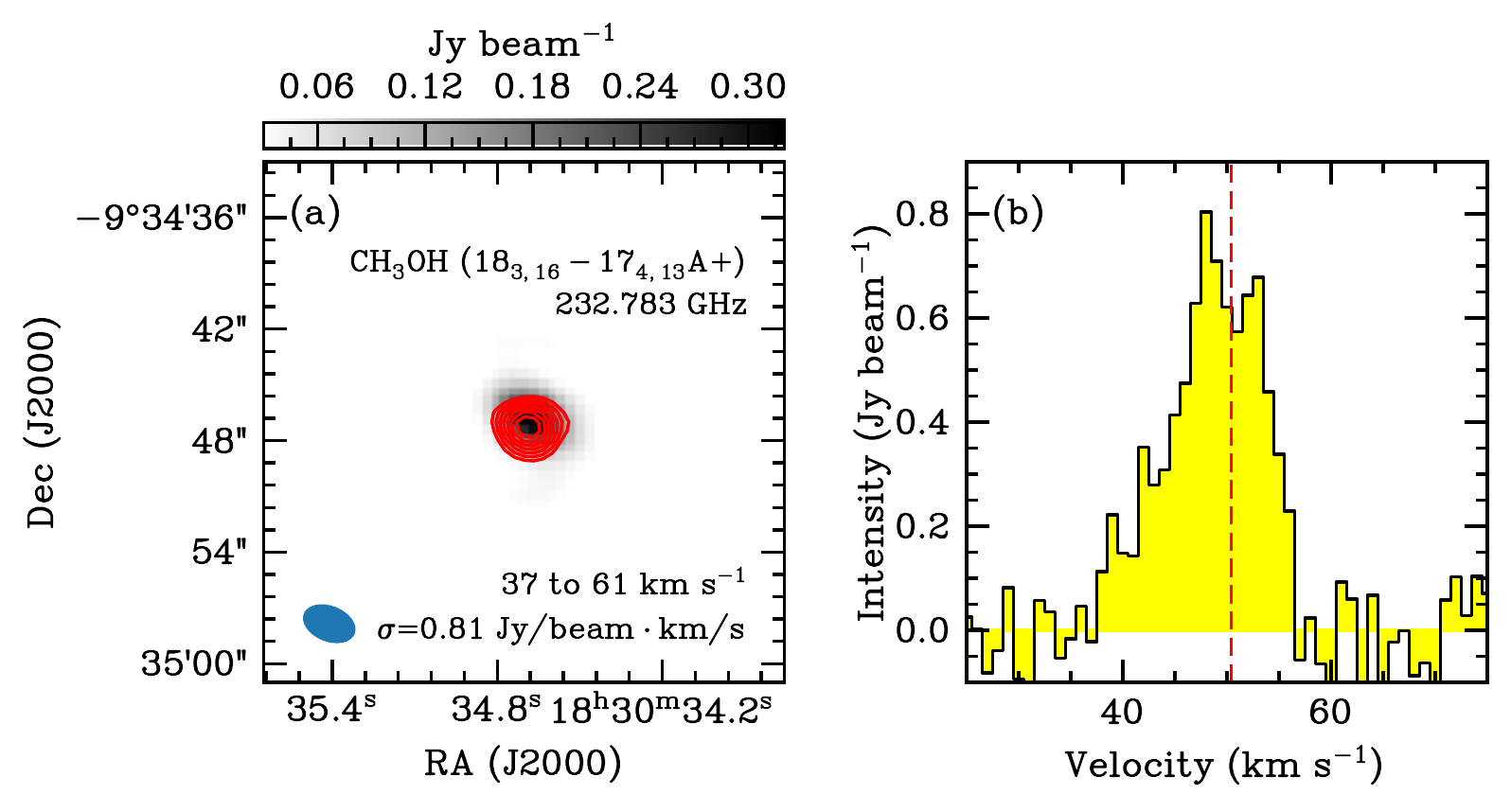}
		\caption{\textit{(a)} Velocity integrated intensity  of
			\methanol~($18_{3,16}-17_{4,13}$A+)
			overlaid on the 1.3 mm continuum image. 
			Contours start from $3\sigma$ 
			with a increasing step of $1\sigma$.  
			\textit{(b)} Spectrum of \methanol~($18_{3,16}-17_{4,13}$A+)
			at the SMA1 peak. The vertical dashed line marks
			the systemic velocity (50.4 \kms) of SMA1. 
			\label{fig-meth-infall}}
	\end{figure}

    Near the peak of SMA1, there is a mid-IR point source (SSTGLMC G022.0387+0.2222, 
    MIR1 for short).
    Photometric data collected from the GLIMPSE, MIPSGAL, Hi-GAL, and ATLASGAL surveys
    have been fitted to models of YSOs 
    developed by \citet{2006ApJS..167..256R,2007ApJS..169..328R}.
    The resultant spectral energy distribution of MIR1 is shown in Figure \ref{fig-sed}.
    The good fits suggest that MIR1 is a Stage 0/I type high-mass prototar with 
    a stellar mass of $[11-15]$ \msun, a total luminosity of $[3.5-7.2]\times10^{3}$ 
    \lsun, an envelope accretion rate of $ [1.3-4.1]\times10^{-3} $ \msun~yr$^{-1}$, 
    and an age of $[0.6-3.8]\times10^{4}$ yr.
    
    \subsubsection{Possible Core-fed Accretion}
    
    As shown in Figure \ref{fig-infall-tracers} (b),
    blue profile and inverse P-cygni profile have been 
    observed in SMA \tco~$ (2-1) $ and
    \ceo~$ (2-1) $ spectra. A similar blue-skewed profile is 
    also seen in \methanol~($18_{3,16}-17_{4,13}$A+) 
    (see Figure \ref{fig-meth-infall}). Although the self-absorption in these lines 
    can be partially due to the short spacing issues related to interferometer observations,
    blue-skewed profiles still could be tracing some dynamical motions. Additionally,
    the inverse P-cygni profile of  \ceo~$(3-2)$ cannot be fully explained by
    short spacing problem, and has been frequently interpreted as a 
    convincing tracer of infall motions in other interferometer observations
    \citep{2009ApJ...697L.116W,2011ApJ...728...91L,2011ApJ...730..102L,
    	2013MNRAS.436.1335L,
    	2011ApJ...728....6Q,2012ApJ...756..170Q}.
    And the short spacing problem would filter 
    structures larger than 20\arcsec and cannot significantly alter the profile
    of \methanol~($18_{3,16}-17_{4,13}$A+) as this line mainly traces the dense core
    (see Figure \ref{fig-meth-infall} (a)). Careful inspection of moment 1 maps of all
    transitions shows there is no detectable rotation in SMA1 with current SMA
    observations. This suggests that the blue-skewed profiles could be due to infall
    motions. Blue profiles of \tco~$ J=2-1 $ based on SMA observations 
    have also been interpreted as tracing infall in other star-forming regions 
    \citep[e.g.,][]{2006ApJ...639..292L,2009ApJ...697L.116W,2014ApJ...784..107Z}.
    
    Using the model of \citet{1996ApJ...465L.133M}, an infall velocity of 
    about 0.16 \kms~was estimated. Assuming a density profile of
    $ \rho\propto r^{-1.5} $, we obtain a mass infall rate of
    $ 7.0\times10^{-5} $ \msun~yr$^{-1}$ {which is a quarter of the
    free-fall accretion rate 
    ($\dot{M}_\mathrm{ff} \sim3\times10^{-4} $ \msun~yr$^{-1}$) and} 
	comparable to
    mass accretion rates of high-mass star-forming cores in
    numerical studies 
    \citep{2007ApJ...666..976K,2010ApJ...722.1556K,2011ApJ...732...20K} 
    as well as in interferometric observations 
    \citep[e.g.,][]{2013MNRAS.436.1335L,2014ApJ...791..123W}.
    Assuming that about half of the gas flow can be successfully 
    channeled onto the central protostar, MIR1 would 
    grow to an O9 star in about $ 1\times10^{5} $ yr if 
    such accretion rate is sustained. 
    
    \vskip 0.2cm
    In summary, the gas in G22 is being channeled into the central hub region through
    global filamentary collapse. The most massive clump C1 at the center is 
    collapsing and feeding the embedded hot molecular core SMA1. The high-mass
    protostar MIR1 in SMA1 is also gaining mass from SMA1 where infall motions
    have been tentatively detected.
    
    \subsection{Influence from the Adjacent Bubble} \label{sec-discussion-bubble}
    
    The infrared bubble MWP1G022027+0022159 has been observed in the 
    south of G22 (see Figures \ref{fig-mul-morph} and \ref{fig-Nh2-Tdust} (b)). The
    dark lane observed to the near-IR suggests the bubble is behind G22. This is also 
    supported by the changes of velocities from red to blue for masers associated
    with the outflow lobe R1 (Figure \ref{fig-outflow}). The reversal of maser velocities
    and large intensity gradients of R1 suggest that the bubble might be strongly interacting
    with G22 and modifying the direction of the outflowing gas. 
    
    As the bubble is slightly behind G22, its expansion is apt to push the 
    gas in G22 towards us. And the interacting region would have a higher temperature 
    and more blue-shifted velocities. Thus, any existing self-absorption due
    to the bubble would result
    in line profiles with a higher red peak (i.e., red profiles). The ubiquitous
    blue profiles in C1 cannot originate from the the influence of the IR bubble.
    However, an expansion of the bubble would introduce a blue wing to the spectrum
    of \hcop~$ J=3-2 $ as shown in Figures 
    \ref{fig-infall-tracers} and \ref{fig-infall-model}.

\section{Summary}\label{sec-summary}

	We have presented a detailed study of a hub-filament system in the G22 cloud. 
	The main findings
	of this work are summarized as follows.
	
	\begin{enumerate}
		\item The G22 cloud is composed of four supercritical filaments
		with mass per unit length ranging from {54 to 220} \msun~pc$^{-1} $.
		Velocity gradients along three filaments revealed by \tco~$ J=1-0 $ indicates
		they are collapsing and channeling gas towards the junction. A total mass infall
		rate of {440} \msun~Myr$^{-1}$ suggests that the hub mass would be doubled in
		{6} free-fall times. Although such a high accretion may not be 
		sustained for almost {two} million years, it is 
		tenable to posit that a fraction of the mass in the central 
		clumps C1 and C2 have been built through 
		large-scale filamentary collapse.
		
		\item The most massive and densest clump C1 at the junction is globally 
		collapsing. With a virial parameter $ \alpha_\mathrm{vir}<1 $, C1 is supercritical
		unless strong magnetic fields can provide additional support. Prevalent 
		blue-profiles of \hcop~$ (3-2) $ and \tco~$ (3-2) $ spectra support a scenario of
		clump-scale collapse. An infall velocity of 
		{0.31} \kms~was estimated via modeling
		the clump-averaged \hcop~spectrum using the RATRAN code. The inferred mass
		infall rate is about $ 7.2\times10^{-4} $ \msun~yr$^{-1}$.
		
		\item Embedded in clump C1, a hot molecular core (SMA1) has been revealed by the SMA 
		observations. More than 100 lines from 19 species (26 isotopologues including
		complex organic molecules) have been detected. The gas temperature was estimated
		to be higher than 200 K via fitting the rotation diagrams of \methanol~and \ace.
		The detected outflows and the Class-I like SED of the embedded mid-infrared source
		suggest that high-mass star formation is taking place in SMA1.
		The high-mass star-forming hot core SMA1 may be still growing
		in mass via clump-fed accretion. This is supported by the high mass infall
		rate of C1 which is significantly larger than the mass loss rate of the 
		outflows.
		
		\item A high-mass protostar (MIR1) is located at the center of SMA1 
		and could be the driving source of the observed outflows. 
		The mass of MIR1 is estimated to be in the range 
		[$ 11-15 $] \msun~and may be
		still growing via core-fed accretion which is 
		supported by the detected infall features in SMA1.
		
		\item  Anomalous methanol emission of \methanol~($8_{-1,8}-7_{0,7}$) 
		at 229.759 GHz and \methanol~($4_{2,2}-3_{1,2}$) at 218.440 GHz 
		was detected in three positions around SMA1. Large 
		$8_{-1,8}-7_{0,7}/3_{-2,2}-4_{-1,4}$ and 
		$4_{2,2}-3_{1,2}/3_{-2,2}-4_{-1,4}$ ratios suggest maser emission of these two
		methanol transitions.
		
	\end{enumerate}

	The coexistence of infall through filaments, clumps and the
	central core has revealed continuous mass growth from large to 
	small scales. This suggest that pre-assembled mass reservoirs may not be 
	indispensable to form high-mass stars. In the process of high-mass
	star formation, the masses of the central protostar, the core, and
	the clump can simultaneously grow via core-fed/disk accretion, 
	clump-fed accretion and, filamentary/global collapse.
	
\begin{acknowledgements}
   	
   	We are grateful to the anonymous referee for the 
   	constructive comments that helped us improve this paper.
   	Prof. Neal J. Evans II and Dr. Kee-Tae Kim are also acknowledged
   	for useful discussions.
   	This work is supported by the National Natural Science 
   	Foundation of China through grants of 11503035, 11573036, 
   	11373009, 11433008 and 11403040, 11403041, the International 
   	S\&T Cooperation Program of China through the grant of 
   	2010DFA02710. Tie Liu is supported by the EACOA fellowship.
   	KW is supported by grant 
   	WA3628-1/1 through the DFG priority program 1573 
   	``Physics of the Interstellar Medium''.
   
%

\end{acknowledgements}

\bibliography{g22}

\appendix

\begin{figure*}[htb]
	\vspace{0.3cm}
	\centering
	\includegraphics[width=0.95\textwidth]{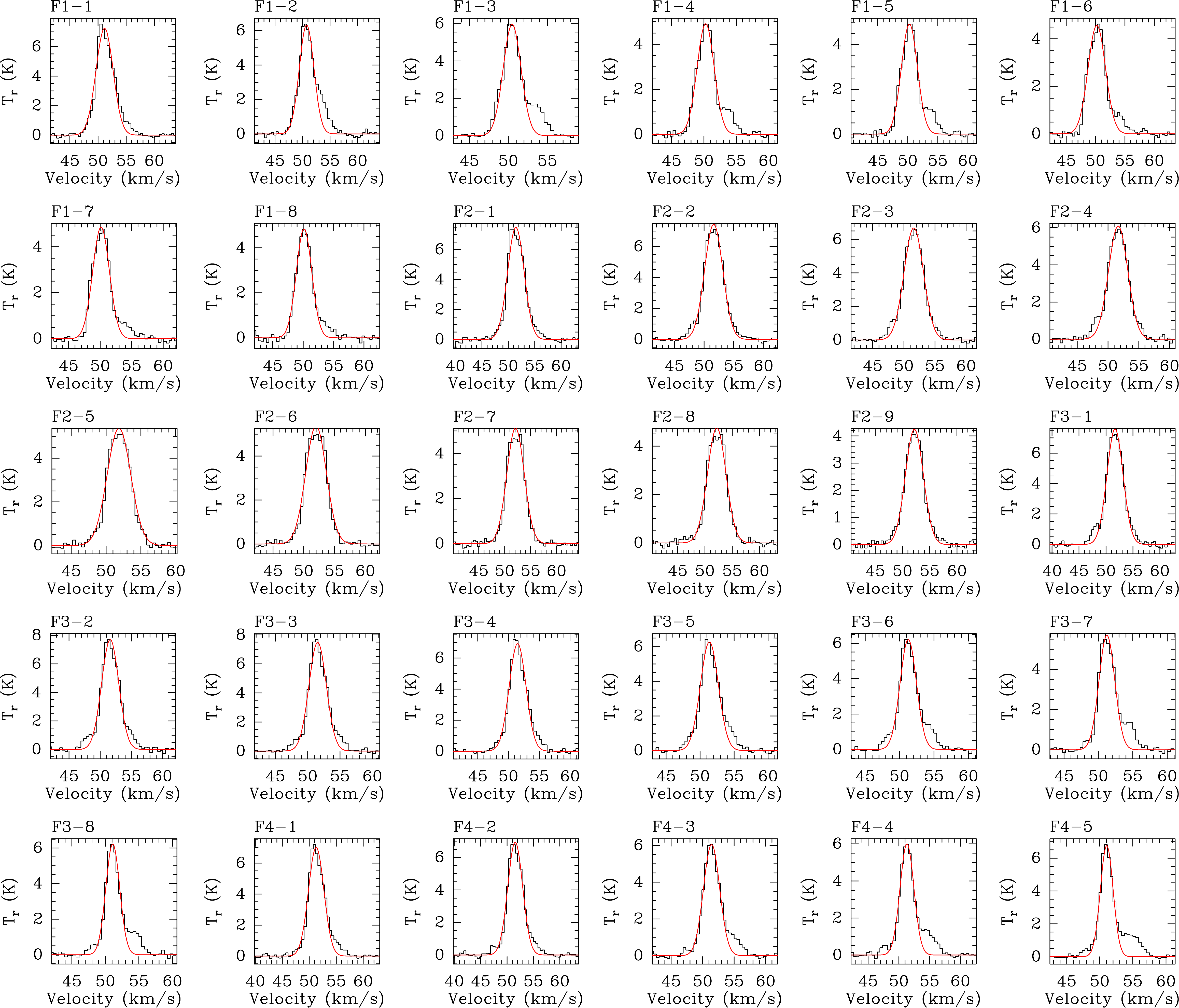}
	\caption{Spectra of \tco~1-0 along filaments.The red line in
		each panel shows the one-dimensional Gaussian fitting result.
		\label{fig-tcoSpec}}
\end{figure*}
\end{document}